\documentclass[twocolumn]{aastex631} 
\usepackage{amssymb,amsmath,graphicx,natbib,hyperref}
\usepackage{multirow}
\usepackage{subfigure}
\usepackage{float}
\usepackage{footmisc}
\usepackage{color}
\newcommand{\amin}{$^{\prime}$}
\newcommand{\asec}{$^{\prime\prime}$}
\newcommand{\kms}{\,km\,s$^{-1}$}
\newcommand{\ms}{\,m\,s$^{-1}$}
\newcommand{\um}{\,$\mu$m}
\newcommand{\pc}{$\%$}

\newcommand \percmsq {\,cm$^{-2}$}
\newcommand \percmcu {\,cm$^{-3}$}

\newcommand \co {$^{12}$CO}
\newcommand \thco {$^{13}$CO}
\newcommand \ceio {C$^{18}$O}
\newcommand \hcn {HCN}

\newcommand \hcop {HCO$^+$}

\newcommand{\tas}		  {$T_a^*$}
\newcommand{\tex}		  {$T_\mathrm{ex}$}

\newcommand{\tr}		  {$T_\mathrm{R}$}
\newcommand{\tmb}		  {\mbox{$T_{MB}$}}
\newcommand{\nhtwo} {$n_{\textrm{H}_2}$}
\newcommand{\ndvv}		  {$N_{\textrm{H}_2}/dv$}
\newcommand{\tauu}		  {$\tau$}

\shortauthors{Li et al.}

\usepackage{enumitem}
\setlist[itemize]{noitemsep, topsep=0pt}
\begin{document}

\title{GBT/\textit{Argus} Observations of Molecular Gas in the Inner Regions of IC\,342}

\shorttitle{\textit{Argus} Observations of IC\,342}
\shortauthors{Li et al.}

\correspondingauthor{Jialu Li}
\email{jialu@astro.umd.edu}

\author[0000-0003-0665-6505]{Jialu Li}
\affiliation{Department of Astronomy, University of Maryland, College Park, MD 20742, USA}

\author[0000-0001-6159-9174]{Andrew I. Harris}
\affiliation{Department of Astronomy, University of Maryland, College Park, MD 20742, USA}

\author[0000-0002-5204-2259]{Erik Rosolowsky}
\affiliation{Department of Physics, University of Alberta, Edmonton, AB, Canada}

\author[0000-0002-3227-4917]{Amanda A. Kepley}
\affiliation{National Radio Astronomy Observatory, 520 Edgemont Road, Charlottesville, VA 22903, USA}

\author[0000-0003-1924-1122]{David Frayer}
\affiliation{Green Bank Observatory, Green Bank, WV 24944, USA}

\author[0000-0002-5480-5686]{Alberto D. Bolatto}
\affiliation{Department of Astronomy, University of Maryland, College Park, MD 20742, USA}

\author[0000-0002-5480-5686]{Adam K. Leroy}
\affiliation{Department of Astronomy, The Ohio State University, 140 West 18th Avenue, Columbus, OH 43210, USA}

\author[0000-0002-3106-7676]{Jennifer Donovan Meyer}
\affiliation{National Radio Astronomy Observatory, 520 Edgemont Road, Charlottesville, VA 22903, USA}

\author{Sarah Church}
\affiliation{Physics Department, Stanford University, Stanford, CA 94305, USA}

\author[0000-0002-7524-4355]{Joshua Ott Gundersen}
\affiliation{Department of Physics, University of Miami, 1320 Campo Sano Avenue, Coral Gables, FL 33146, USA}

\author[0000-0002-8214-8265]{Kieran Cleary}
\affiliation{California Institute of Technology, 1200 E. California Blvd., Pasadena, CA 91125, USA}

\author{DEGAS team members}

\begin{abstract}

We report observations of the ground state transitions of \co, \thco, \ceio, HCN, and \hcop\ at 88--115~GHz in the \textcolor{black}{inner region} of the nearby galaxy IC\,342. {These data} were obtained with the 16-pixel spectroscopic focal plane array \textit{Argus} on the 100-m Robert C. Byrd Green Bank Telescope (GBT) at 6--9\asec\ resolution. In the nuclear bar region, the intensity distributions of \co(1--0) and \thco(1--0) emission trace moderate densities, and differ from the dense gas distributions sampled in \ceio(1--0), HCN(1--0), and \hcop(1--0). 
We observe a constant HCN(1--0)-to-\hcop(1--0) ratio  of 1.2$\pm$0.1 across the whole $\sim$1~kpc bar. {This indicates that HCN(1--0) and \hcop(1--0) lines have intermediate optical depth, and that the corresponding \nhtwo\ of the gas producing the emission is of 10$^{4.5-6}$~\percmcu.} We show that \hcop(1--0) is thermalized and HCN(1--0) is close to thermalization. The very tight correlation between HCN(1--0) and \hcop(1--0) intensities across the 1~kpc bar suggests that this ratio is more sensitive to the relative abundance of the two species than to the gas density. We confirm the angular offset ($\sim$10\asec) between the spatial distribution of molecular gas and the star formation sites. {Finally, we find a breakdown of the $L_\textrm{IR}$-$L_\textrm{HCN}$ correlation at high spatial resolution due to the effect of incomplete sampling of star-forming regions {by HCN emission} in IC\,342. The scatter of the $L_\textrm{IR}$-$L_\textrm{HCN}$ relation decreases as the spatial scale increases from 10\asec~to 30\asec~(170--510~pc), and is comparable to the scatter of the global relation at the scale of 340~pc.  }

\end{abstract}
    
\section{Introduction}

Understanding the distribution, physical conditions, and dynamics of molecular gas is essential to understanding star formation. Studies
of individual regions in our galaxy have shown that stars form in dense molecular clouds \citep[e.g., see][]{lada03, heiderman10, lada10}, while \textcolor{black}{resolved} observations of external galaxies provide global information related to star formation efficiencies within nuclei, spiral arms, and other regions \citep[e.g., see][]{silk97, kennicutt98, elmegreen02, bigiel08}.  In all cases,
probing the complexity and scale of star formation regions requires a large spatial dynamic range to follow core formation and collapse to circumstellar
scales. 

Rotational transitions of molecules such as \co, HCN, \hcop, and their isotopologues are common tools for probing cold molecular clouds in external galaxies because of their relatively high abundance and the excitation energy levels from a few to several tens of {kelvins}. Specifically, $J$=1--0 lines of \co, \thco, and \ceio\ trace gas with densities above \mbox{$\sim10^2$--10$^3$~\percmsq}, capturing the bulk of the molecular gas. 
\hcn\ and \hcop\ have higher dipole moments (2.98 and 3.92~D versus 0.11~D for \co), and their ground state transitions trace denser gas (\mbox{$\sim10^4$--10$^5$~\percmsq}). 
{Because of the different critical densities, this set of molecular lines} can establish the quantitative link between gas density and star formation across a variety of environments (e.g., the PHANGS/CO survey, \citealt{leroy21};  the EMPIRE/HCN survey, \citealt{jimnez19}; the DEGAS survey, {Kep ley et al., in prep})

Single transitions cannot be easily used to quantify the exact gas density or temperature, however. The interpretation of the line intensity involves a high degree of degeneracy, including the observational spatial resolution, the optical depth effects, and/or the actual excitation condition, etc. As spatially resolved observations become more accessible, multi-transition ratios, on the other hand, may further {break such a degeneracy} by comparing the observations and theoretical predictions within specific parameter space due to the different excitation properties among different species  {\citep[e.g.,][]{meijerink07, viti17, leroy17}}.

\begin{deluxetable*}{lccc}
\label{tab:obs}
\tablecolumns{4}
\tablecaption{{Summary of Observations}}
\tablehead{ & \co(1--0) & \thco(1--0) $\&$ \ceio(1--0) & HCN(1--0) $\&$ \hcop(1--0) }
\startdata
Observation date & 17-Sep-29 & 17-Oct-18/19, 17-Nov-21 & 17-Oct-25/27, 17-Nov-14\\
Observation ID & GBT17B-412-01/02 & GBT17B-151-01/08 & GBT17B-151-02/03/05 \\
Mapping area & 3.5$'$ $\times$ $5'$& 2.5$'$ $\times$ $2.5'$ & 2.5$'$ $\times$ $2.5'$\\
On-source integration (hrs) & 6.17 & 1.33 & 11.52 \\
Total integration (hrs) & 10.08 & 2.38 & 20.67 \\
Frequency Setup (GHz) & 115.271 & 109.982 & 88.903\\
Beam Size ($''$)& 6.33 & 6.63 & 8.21\\
Beam Size (pc~beam$^{-1}$) & 101& 106& 131\\
Velocity resolution (km\,s$^{-1}$) & 3.8 & 4.0 & 4.9 \\
Main beam efficiency & 28.3\pc & 44.5\pc & 52.5\pc \\
Calibration error\tablenotemark{a} & 25\% & 23\%& 15\% \\
Scan Rate ($''$\,s$^{-1}$) & 1.71 & 1.67 & 1.67\\
{Noise in \tmb~(K)}\tablenotemark{b} & 0.09$\pm$0.02 & 0.018$\pm$0.004 & 0.0035$\pm$0.0005 \\
\enddata
\tablenotetext{a}{Based on GBT memo \#302~\citep{frayer19}.}
\tablenotetext{b}{{The spectral noise level varies at different positions on the map, depending on the integration time within each beam area.}}
\end{deluxetable*}

Among the ratios of transition of different species, the intensity ratio of HCN(1--0)-to-\hcop(1--0), here denoted as $\mathcal{R} \equiv~I$(HCN)/$I$(\hcop), {is of specific interest in this paper}. {The critical density of \hcop(1--0) is lower by 5--20 times than that of HCN(1--0) at temperatures of 20--100 K because of their different collisional de-excitation rate coefficients.} The intensity ratio $\mathcal{R}$ is  a potentially valuable probe of the physical conditions of molecular gas in external galaxies because the correlation between the infrared luminosity $L_{IR}$ and $\mathcal{R}$ was established by single beam observations \citep{gg06}. This correlation was further expected to suggest the importance of AGN (active galactic nucleus) in galaxies with higher infrared luminosities \citep{ima06, ima07}, although recent studies with the 30-m IRAM survey \citep{privon15, privon20} conclude that globally enhanced HCN emission relative to \hcop\ is not correlated with the presence of an AGN.  

The DEGAS (Dense Extragalactic GBT\footnote{The Green Bank Observatory is a facility of the National Science Foundation operated under cooperative agreement by Associated Universities, Inc.}+\textit{ARGUS} Gas Density Survey; Kepley et al., in prep; proposal ID: GBT17B-151, GBT17A-304, GBT17A-212) survey conducts efficient multi-line spectral mapping on the ground transitions of \thco, \ceio, HCN, and \hcop\ of 17 external galaxies (Kepley et al., in prep). Specifically, DEGAS uses the \textit{Argus} focal plane array spectroscopic imager \citep{sieth14} and the spatial resolution (6--8\asec) provided by the 100-m aperture of the GBT from 86--115~GHz. For each galaxy, the central 2$^{\prime}\times2^{\prime}$ region was covered. 

{In this paper, we choose IC 342, the nearest galaxy in the DEGAS sample, as the ideal target for a detailed spatially resolved multi-line study. IC\,342 is a nearly face-on, spiral galaxy at 3.3 Mpc~\citep{saha02} harboring starburst activity in its center. Its large angular size~\citep[20$^\prime$ in optical and 80$^\prime$ in HI,][]{crosthwaite00} is well suited to detailed imaging with the GBT's beam sizes, while the concentration of molecular gas in the galactic center gives rich line emission~\citep[e.g.,][]{downes92, turner92, meier00}. In addition to the ground transitions of \thco, \ceio, HCN, and \hcop\ observed by DEGAS, we obtained observations of \co\ from a DDT (Director’s Discretionary Time) proposal toward the inner disk region (3.5$^{\prime}\times5^{\prime}$) of IC\,342 (proposal ID: GBT17B-412). These single-dish observations provide high spatial resolution without resolving out extended emission, as an interferometer would.}

The structure of this paper is as follows. We summarize our observations and data reduction in Section~\ref{sec:obs}. In Section~\ref{sec:results}, we describe the distribution of molecular material across the nucleus and inner spiral arms in \co, and along the nuclear bar of IC\,342 in all five molecular species. In Section~\ref{sec:discussion}, we discuss the implication of the constant HCN-to-\hcop\ intensity ratio observed across the entire 1~kpc bar of IC\,342, and specifically analyze the breakdown of the $L_\textrm{IR}$-$L_\textrm{HCN}$ correlation of IC\,342 at high spatial resolution.  Section~\ref{sec:summary} is a summary overview, and in the Appendix, we provide some detailed information connected with our analysis.

\section{Observations and Data Reduction} \label{sec:obs}

We used the \textit{Argus} 16-pixel spectroscopic focal plane array on the 100-m diameter Robert C.\ Byrd Telescope of the Green Bank Observatory to conduct observations of IC\,342 from 2017 September to 2017 November. A summary of observational parameters is  in Table~\ref{tab:obs} for the five molecular species we imaged. The on-source time for all species in total was 19~hrs out of the total observing time (including overhead) of 33~hrs (see Table~\ref{tab:obs}). The telescope slewed across a region of the sky in a raster pattern to make OTF {(On-The-Fly)} maps \citep{haslam70, mangum07}, with the data and the antenna position recorded every 0.5~s. Pointing and focus were calibrated every 30--40 minutes with the source 0359+5057. The backend, VEGAS, was configured in mode 1 with a single spectral window with a bandwidth of 1500 MHz (1250~MHz effective) and a spectral resolution of 1465 kHz. {\thco(1--0) and \ceio(1--0) were observed simultaneously in a spectral window centered at 109.99176~GHz, and HCN(1--0) and \hcop(1--0) were observed simultaneously in a spectral window centered at 88.91123~GHz.} Observations were mostly conducted under windless conditions ($v_\textrm{rms}<$~1\ms). For \co, however, for half of the observing time, $v_\textrm{rms}$ was greater than 5\ms, which caused pointing deviations as high as $\sim$~1 beam ($\sim$~6\asec\ at 115~GHz). We established pointing corrections by cross-correlating each 40~min individual observing session with \co\ maps observed {under low wind speed}.  

We calibrated and post-processed data with the Python packages \texttt{gbtpipe} and \texttt{degas}. {Initially, \textit{Argus} data are  on \tas\ scale} after  chopper wheel calibration \citep{kutner81}. We obtained atmospheric temperature and  opacity information from the GBT weather database and then corrected for atmospheric attenuation, resistive losses, rearward spillover, and scattering.  Our final brightness temperature scales are in $T_\mathrm{MB}$ to better represent the intensity in the main beam. We used conversion factors between \tas\ and $T_\mathrm{MB}$ of $\eta_\textrm{MB}$ of 28.3\pc\ for \co, 44.5\pc\ for \thco\ and \ceio, and 52.5\pc\ for HCN and \hcop~\citep[GBT Memo~\#302,][]{frayer19}. Next, problematic spectra were automatically detected and dropped. {The reduction pipeline flagged and dropped  spectra with spikes or an RMS noise level higher than 1.1 times expected from the radiometer equation.} We then removed low-order polynomial baselines and interpolated them to a regular grid in data cubes. We used the \co\ map as a spatial and velocity space mask to constrain the emission regions for the other molecules. We integrated the data cubes from $-100$ to $150$~\kms\ to make moment maps using functions from the \texttt{SpectralCube} package. 

\section{Results}\label{sec:results}

\subsection{Spatial Distribution of the Molecular Gas}\label{subsec:31}

\begin{figure*}
    \centering
    \includegraphics[width=\linewidth]{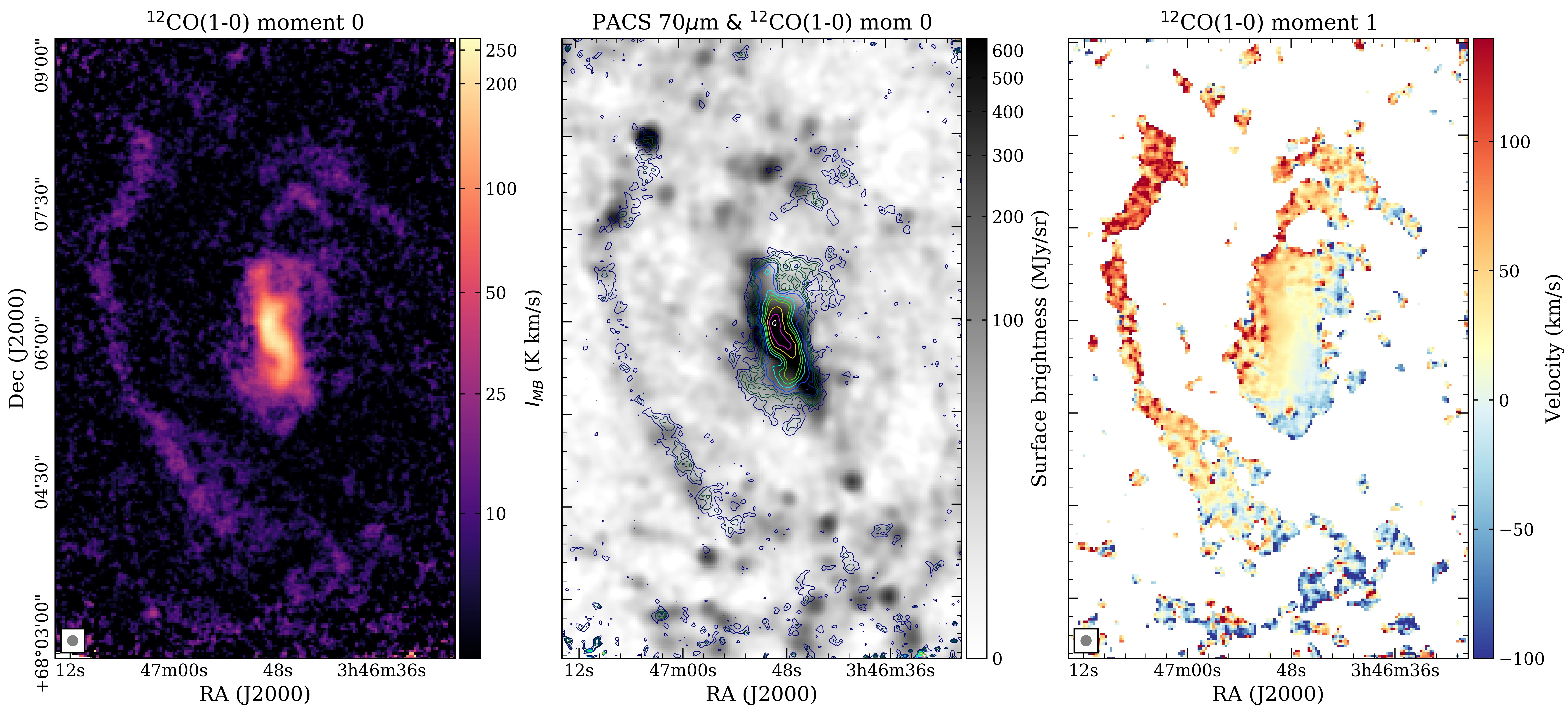}
    \caption{\textit{Left:} the GBT/\textit{Argus} \co~(1--0) moment 0 image with an arcsinh intensity stretch. The 6.3\asec\ beam is in the box at the lower left of the plot. The 1$\sigma$ level of the integrated intensity is 2.8~K\kms. \textit{Middle:} \co~(1--0) moment 0 contours overlaid on the \textit{Herschel} PACS 70~\um\ image \citep{kennicutt11}, indicating correspondence between the molecular gas distribution and the star formation locations.  The contour levels of \co\ are 10, 15, 20, 25, 30, 40, 60, 80, 100, 150, 200, 250~K~\kms. We note that due to the large dynamic range of the \textit{Herschel} map, the surface brightness in the central nuclear region (radius $\sim$10\asec) is much brighter than the upper limit of the color bar and can be as high as 20,000~MJy/sr. \textit{Right:} the intensity-weighted mean velocity map of \co.} 
    \label{fig:12co}
\end{figure*}

\begin{figure*}[!t]
    \centering
    \includegraphics[width=\linewidth]{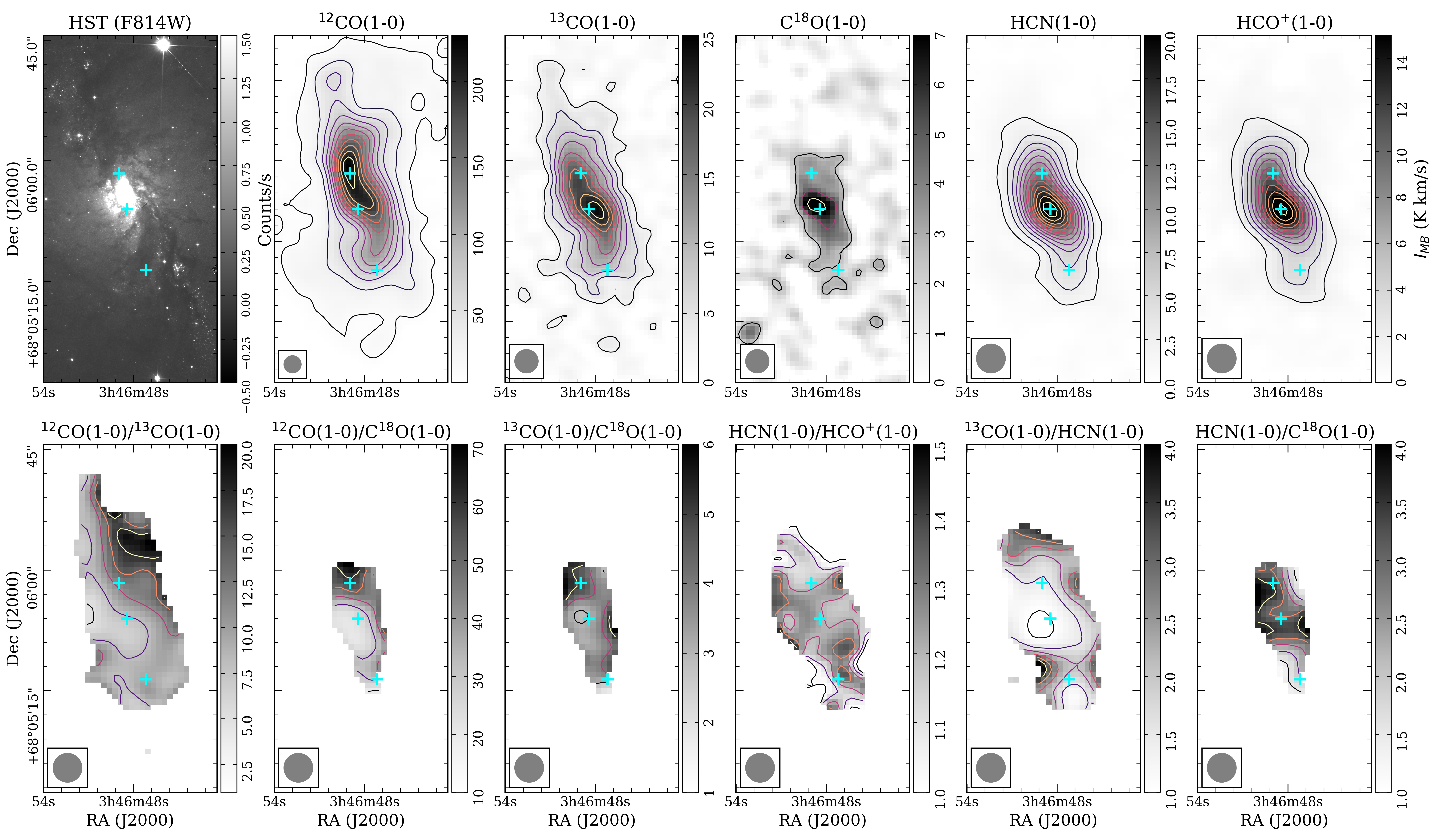}
    \caption{\textit{Upper panels}: \textit{HST} (F814W) map (\textit{lower left}) and the integrated intensity maps of \co, \thco, \ceio, HCN, \hcop(1--0) in the central bar region in beam sizes of 8.2\asec, 8.6\asec, 8.6\asec, 10.7\asec, and 10.7\asec. Contours start from 10$\sigma$, 1$\sigma$, 1$\sigma$, 3$\sigma$, 3$\sigma$ and are in steps of 20$\sigma$, 1$\sigma$, 1$\sigma$, 3$\sigma$, 3$\sigma$. The correspondent 1$\sigma$ levels are 1, 2.5, 2.5, 0.5, and 0.5~K~\kms\ for all the five species. The beam resolutions are increased by 1.3 times for consistency with the DEGAS survey for better SNRs. Spectra measured at the positions of the three cyan crosses within a 10.7\asec\ beam are plotted in Figure~\ref{fig:red} for a comparison of the line properties. {\textit{Lower panels}: ratio maps of the integrated intensities above 3$\sigma$ between different molecular lines. All maps are convolved to the same resolution of 10.7\asec\ before the ratio maps are made. Contours of the ratio start from 5, 10, 3, 1, 1, 1 from left to right and are in steps of 3, 30, 1, 0.1, 0.5, 0.5. Scatter plots comparing line intensities pixel-by-pixel are presented in Figure~\ref{fig:ratio}.}}
    \label{fig:all}
\end{figure*}

\begin{figure*}
    \centering
    \includegraphics[width=\linewidth]{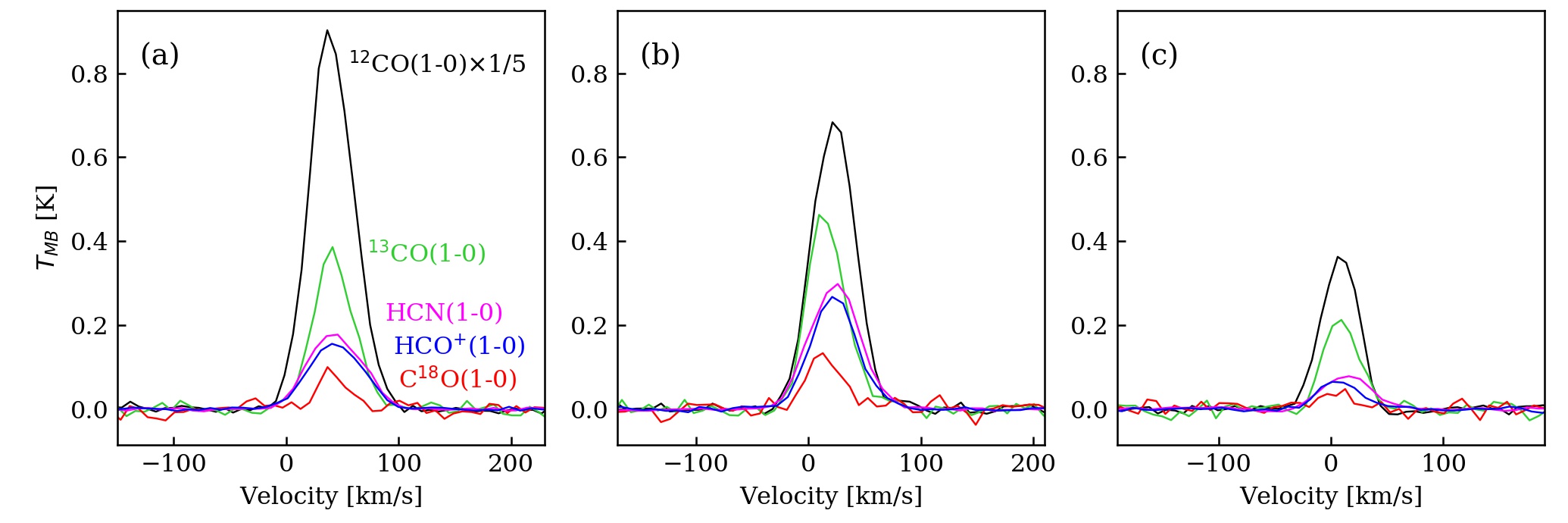}
    \caption{Spectra of \co(1--0) (scaled by 1/5), \thco(1--0), \ceio(1--0), HCN(1--0), and \hcop(1--0) from the three 10.7\asec\ beams whose centers are marked as cyan crosses in Figure~\ref{fig:all} (from top to bottom), representing that the line as well as the velocity-integrated intensity of all species other than \co\ peak at the nuclear center and decrease towards the bar ends. Shifts of the enters of the emission lines indicate the kinematic rotation of the molecular bar.}
    \label{fig:red}
\end{figure*}

\begin{figure*}[!t]
\begin{center}
\includegraphics[width=0.98\textwidth]{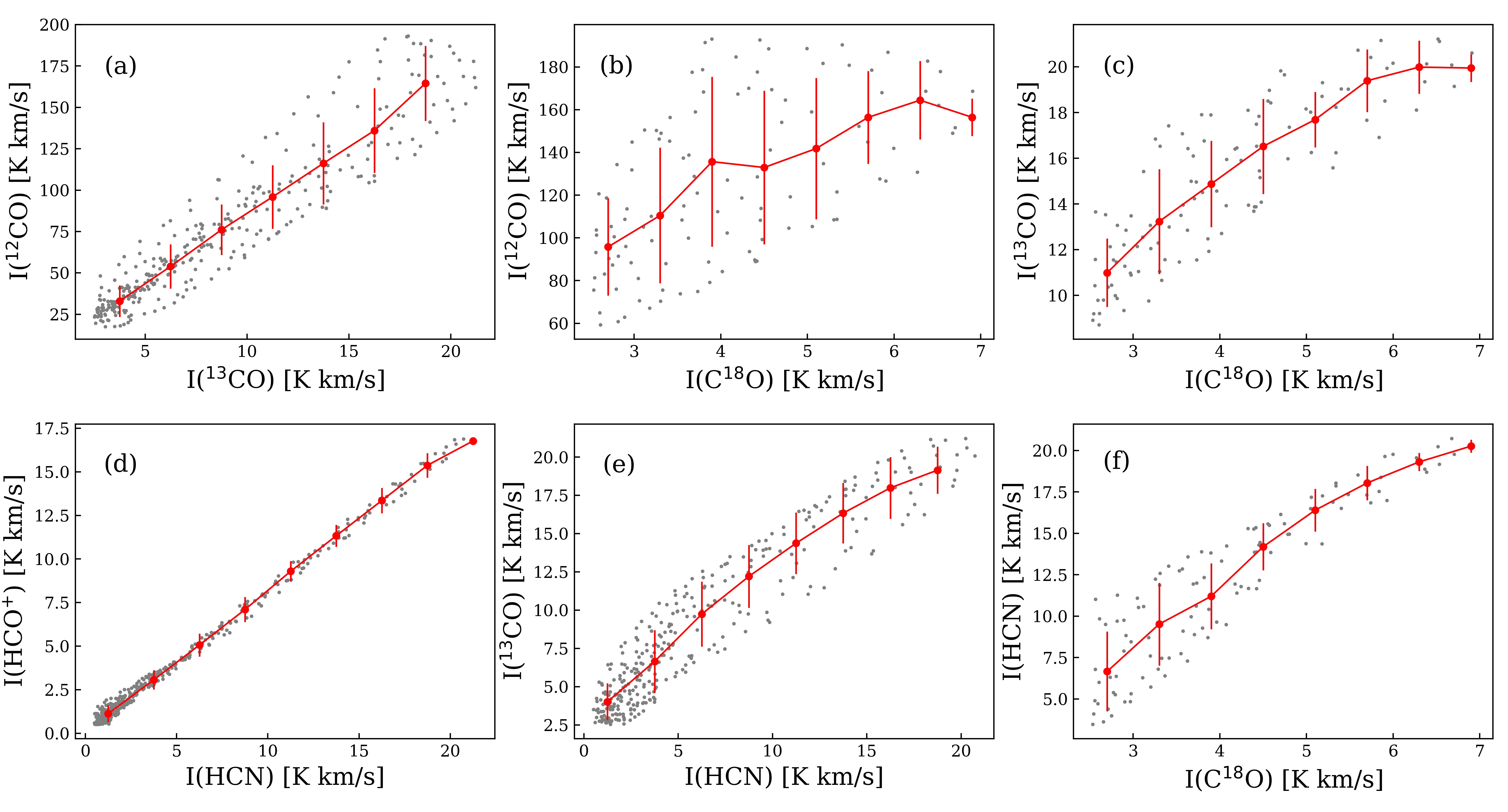}
\caption{A comparison of the integrated intensities of (a) \co/\thco, (b) \co/\ceio, (c) \thco/\ceio, (d) \hcop/HCN, (e) \thco/HCN, and (f) \ceio/HCN from individual image pixels from a 0.9\amin$\times$1.7\amin\ region that covers the galactic bar. Maps of all the five molecules are convolved to the same resolution of 10.7\asec.  \label{fig:ratio}
}
\end{center}
\end{figure*}

We present the moment 0 (integrated intensity) and the moment 1 (intensity-weighted mean velocity) maps of \co(1--0) from the inner disk region of IC\,342 in Figure~\ref{fig:12co}. The size of the maps is 3.5\amin$\times$5\amin. In the middle panel of Figure~\ref{fig:12co}, the \co\ moment 0 map is overlaid in contours on the \textit{Herschel} PACS 70~\um\ map for a comparison with the distribution of dust heated by young stars. 

Ground transitions of \thco, \ceio, HCN, and \hcop\ were observed in a smaller region in size of 2.5\amin$\times$2.5\amin\ and only probe the bar region of IC\,342. {The calibration error is 25\pc\ for \co, 23\pc\ for \thco\ and \ceio, and 15\pc\ for HCN and \hcop~\citep[GBT Memo~\#302,][]{frayer19}.
Comparing against \co, the relative calibration error uncertainty {in ratios is} $\sim$30\% between nights.} The integrated-intensity images of the \mbox{$\sim$1.4\amin\ (1.3~kpc)} long bar region in all five molecular species, including \co, are presented in Figure~\ref{fig:all}. For a better signal-to-noise ratio, all images other than that of \co\ are spatially smoothed to an output resolution 1.3 times larger than the input resolution following the procedure of the DEGAS survey (Kepley et al., in prep). 

Figures~\ref{fig:12co} and \ref{fig:all} show that the spatial distributions of these molecular gas species closely match past observations \citep{downes92, turner92, meier00, mt01, kuno07, hirota10}. The large scale \mbox{\co(1--0)} image reveals a morphology with two asymmetrical arms extending from the ends of a nuclear bar oriented roughly from north to south. \co\ emission peaks at $\sim$0.2\amin\ (191~pc) north of the nuclear center, while the peak brightness and the integrated intensities of the other four species all peak at the nuclear center and decrease towards the bar ends (see Figures~\ref{fig:all} and \ref{fig:red}). Among these species, the morphology of \thco\ most resembles \co, with a rather elongated structure and two tentacle-like substructures to the north. {However, unlike \co, the intensity peak of \thco\ is located at the nuclear center.} \ceio, HCN, and \hcop\ have more concentrated structures{ and illustrate the ``S-shape" shown in the central region of \co\ and \thco.} The morphology of HCN and \hcop\ emission is essentially identical (Figure~\ref{fig:all}).  

We compare the integrated intensities measured by GBT/\textit{Argus} to those observed by other single-dish observations. For \co, the peak values of the integrated intensities on the image observed by the 45-m telescope \citep[15\asec,][]{kuno07} and the GBT/\textit{Argus} are 180 and 257~\mbox{K\,\kms}. Convolving the \textit{Argus} map to a 15\asec\ beam results in a peak \co\ intensity of 162~\mbox{K\,\kms}. For \thco, the values are 22~\citep[20.4\asec,][]{hirota10} and 24~\mbox{K\,\kms}. The peak \thco\ value from the convolved \textit{Argus} map is 14~\mbox{K\,\kms}. For HCN and \hcop, we compare the results from the 30-m telescope \citep[$\sim$26\asec,][]{ng92}, {for which the two lines were observed separately}. Values of HCN are 19.1 (30-m telescope), 21 (data from this paper, before convolution), and 8.5 (data from this paper, after convolution)~\mbox{K\,\kms}, separately. Values of \hcop\ are 9.8, 17, and 7.0~\mbox{K\,\kms}. While the HCN intensity measured by the 30-m telescope is about twice the \hcop\ intensity (19.1 vs. 9.8~\mbox{K\,\kms}), the two are similar (21 vs. 17~\mbox{K\,\kms}) in observations of \textit{Argus}. {The HCN and \hcop\ observations presented in this paper have excellent relative calibration because the two species were observed simultaneously in the same sideband and were reduced with the same procedure. Both lines, therefore, share pointing and focus errors and changes in atmospheric transmission, permitting accurate comparisons of relative intensities and spatial distributions.}

{To compare the distribution of molecular gas to dust-reprocessed UV radiation from young stars,} we overlaid the contours of the integrated intensity of CO on a \textit{Herschel} 70~\um\ image \citep{kennicutt11} in the middle panel of Figure~\ref{fig:12co}. Overall, within the inner disk region of IC\,342, the brightest \co\ emission follows the infrared emission. In the southern and the northern arms, however, peaks of \co\ intensity do not coincide with peaks of the infrared emission. A similar spatial offset is also shown in the nuclear bar region, as most \co\ emission is located at the concave side of the ``S-shape" infrared bar.

The spatial offset seen between \co~(and/or \thco) and 70~\um\ emission in the bar region is consistent with the spatial offset seen between \thco\ and H$\alpha$ emission \citep[5--10\asec, ][]{turner92} in either the size and the direction. While H$\alpha$ emission traces young stars, it suffers from extinction. The consistency between IR and H$\alpha$ emission, therefore, implies that young stars emitting in H$\alpha$ are not entirely hidden in the dust. $HST$ archival images (PID 5446, 6367, see the lower left panel in Figure~\ref{fig:all}) also confirm that in the bar region, young stars emit through the mixed filamentary dusty structures. Therefore, the angular offset between the \co~{(and/or \thco)} emission region and the star formation sites appears to be a physical separation rather than caused by variable dust extinction.

\subsection{Line Intensity Ratios of Different Molecular Species in the Nuclear Bar Region}\label{subsec: 32}

We compare the intensities of the lines of the five molecular species to understand their ability to probe different physical conditions in the nuclear bar region of IC\,342. To quantify the differences shown in molecular emission distributions, we convolved all images to 10.7\asec\ (the resolution of the HCN and \hcop\ maps; see \S~\ref{subsec:31}) resolution. {We present the ratio maps in Figure~\ref{fig:all}}, and compared integrated intensities pixel-by-pixel between different species in scatter plots in Figure~\ref{fig:ratio}. {The ratio maps present how the relative spatial distributions change, while the scatter plots provide information of the overall changing trend as the line intensities change.}

The interpretation of the physical conditions of the line ratios is considered from two aspects: the optical depth and the excitation condition. Because all five species have similar line widths (see Figure~\ref{fig:red}), the integrated intensity ratio between any two lines, or $\mathcal{R}$, is mainly determined by the peak temperatures of the two line spectra. For each individual line, the peak temperature is
\begin{equation}
    T_R = J_\nu(T_\textrm{ex})  \left (1-e^{-\tau_\nu} \right)f_c,\label{eqn:1}
\end{equation}
where $T_\textrm{ex}$ is the excitation temperature, $\tau$ is the line optical depth, and $f_c$ is the filling factor. {$J(T_\textrm{ex})$ is the Rayleigh-Jeans corrected excitation temperature defined as follows}:
\begin{equation}
    J_\nu(T) \equiv \frac{h\nu/k}{e^{h\nu/kT}-1}.
\end{equation}
When there is no partial coverage ($f_c=1$), $J(T_\textrm{ex})$ approaches {the kinematic temperature, $T_\textrm{kin}$}, if the lines are fully thermally excited, i.e., the volume density is higher than $n_\textrm{crit}$ and the $J$=1--0 transition is collisionally de-excited. {Therefore, if both lines are thermalized, the line intensity ratio equals the ratio of $(1-e^{-\tau})$.  Specifically, if the lines are optically thin ($1-e^{-\tau}\sim\tau$), we can interpret the line intensity ratio as the column density ratio, or equivalently, the abundance ratio.} Otherwise, if the lines are optically thick, the line intensity ratio would be the ratio of $T_\textrm{ex}$. 

We describe the intensity ratios of different molecular species in IC\,342 indicated by \textit{Argus} (Figure~\ref{fig:ratio}) in the following paragraphs, relying on intuition informed by the RADEX~\citep{vdt07} modeling process described in much more detail in Section~\ref{subsec:ratio}. A summary follows describing the implications for each ratio that we examined, as well as contextual references for the same ratios studied in similar environments:

\paragraph{\co\ vs. \thco} {While the ratio map between \co\ and \thco\ shows that their intensity peaks do not overlap,} Figure~\ref{fig:ratio}a shows that \mbox{$I$($^{12}$CO)} and \mbox{$I$($^{13}$CO)} overall have a linear correlation. \mbox{$I$($^{12}$CO)}/\mbox{$I$($^{13}$CO)} = 10.4 $\pm$ 2.2, and {is within the range of} the ratio \citet{mt01} found, which is from 4 to 12, measured by $\sim4.5$\asec\ resolution by the Owens Valley Radio Observatory Millimeter Interferometer. {The range of \mbox{$I$($^{12}$CO)}/\mbox{$I$($^{13}$CO)}, $\leq10$, is much smaller than the abundance ratio in the Galactic center \citep[24$\pm$7,][]{halfen17}. While this may imply that \thco\ has optical depth $\tau\lesssim1$ while \co\ is optically thick, another possibility is that \thco\ has a smaller areal covering factor relative to \co\ (between $\sim$0.1 to 0.2) because of its substantially lower effective critical density than \co, once radiative trapping is taken into account.} We suggest that \thco\ is not optically thin because of the linear intensity relation with optically thick \co; this conclusion can be also supported by the fact that \thco\ intensity saturates at high \ceio\ intensities (see below).


\paragraph{\co, \thco\ vs. \ceio} {The spatial distribution of \ceio\ is mostly concentrated on the brightest regions of \co\ and \thco\ due to our detection limit on \ceio~(Figure~\ref{fig:all}). The saturation effect is therefore important when comparing the intensities of \ceio\ to \co\ and/or \thco\ {in the brightest region} (Figure~\ref{fig:ratio}b and \ref{fig:ratio}c).} \mbox{$I$($^{12}$CO)}/\mbox{$I$(C$^{18}$O)} and  \mbox{$I$($^{13}$CO)}/\mbox{$I$(C$^{18}$O)} range from 26--42 and from 2.5--4. \mbox{$I$($^{13}$CO)}/\mbox{$I$(C$^{18}$O)} is smaller than a typical Galactic center value of 10 \citep{dahmen98}. {This result suggests that either there is a small filling factor of \ceio\ compared to \thco, or that \thco\ has an intermediate optical depth (see equation~\ref{eqn:1}).} {The saturation of \thco\ at bright \ceio\ suggests that \thco\ with an intermediate optical depth is more likely.} If we assume \ceio\ is optically thin due to its low abundance, we may derive H$_2$ column densities directly from \mbox{$I$(C$^{18}$O)} following equation~2 in \citet{mt01}. For \tex\ between \mbox{10--30~K}, we derive $N(\textrm{H}_2) =$ 1.2--3.1$\times$10$^{22}$~\percmsq. 

\paragraph{HCN vs. \hcop} {The morphology of HCN and \hcop\ emission are almost identical, so the intensity of HCN and \hcop\ has a tight relation: {the Pearson correlation coefficient is 0.998.}} The tight relation yields a line intensity ratio $\mathcal{R}$, {defined as $I$(HCN)/$I$(\hcop),} of $ 1.2\pm0.1$ (Figure~\ref{fig:ratio}d) across the entire 1~kpc bar region. The ratio, which is larger than one, also suggests that both lines likely have intermediate optical depths. {We note that for HCN and \hcop, the critical densities are similar enough that filling factors will also be similar.} Because the interstellar medium and particularly the interstellar medium of a bar region is complex and fractal, it is likely to have gas in a variety of states. {Therefore, such a constancy of $\mathcal{R}$ indicates that $\mathcal{R}$ in the bar region of IC\,342 is insensitive to locally varying physical conditions. HCN(1--0) and \hcop(1--0) lines cannot both be subthermally excited, because emissions from subthermal excitation conditions are strongly influenced by the varying environment across the whole nuclear bar, assuming that the bar is unlikely to be homogeneous for the whole 1~kpc region. We need to look for a parameter space that is {insensitive} to modest changes of physical conditions over a large region and will discuss the certain set of environmental conditions that the constant $\mathcal{R}$ constrains further in \mbox{Sec. \ref{subsec:ratio}}.}

\paragraph{\thco\ and \ceio, vs. HCN and \hcop} We present \mbox{$I$($^{13}$CO)}-$I$(HCN) and $I$(HCN)-\mbox{$I$(C$^{18}$O)} correlations in Figure~\ref{fig:ratio}e and \ref{fig:ratio}f. Only HCN emission is included in the comparison, {because $I$(\hcop) has a tight correlation with $I$(HCN) and nearly identical correlations to intensities of \thco\ and \ceio.} As Figure~\ref{fig:ratio}e and \ref{fig:ratio}f show, {while \mbox{$I$($^{13}$CO)} saturates at high $I$(HCN), $I$(HCN) slightly saturates at high \mbox{$I$(C$^{18}$O)}.} We conclude that the saturation of \mbox{$I$($^{13}$CO)} in high $I$(HCN) is the result of the optical depth effect on \thco, {although we note that HCN approaches large opacity as well at its high intensity. When \thco\ has a high optical depth, it will cover the more extended envelope of the cloud than seen in HCN, which becomes optically thick in the smaller cloud core regions. Compared to \thco, \ceio\ thus is a better tracer of the true distribution of the molecular gas. Compared to HCN and \hcop, \ceio\ constrains the column density of the dense molecular gas in the center of IC\,342 better because its emission is optically thin in that region, but they are comparable at lower intensities.}




\section{Discussion}\label{sec:discussion}

\subsection{$I$(HCN)/$I$(\hcop) As a Relative Abundance Tracer} \label{subsec:ratio}

\begin{figure*}[!t]
\begin{minipage}[t]{\textwidth}
\includegraphics[width=\textwidth]{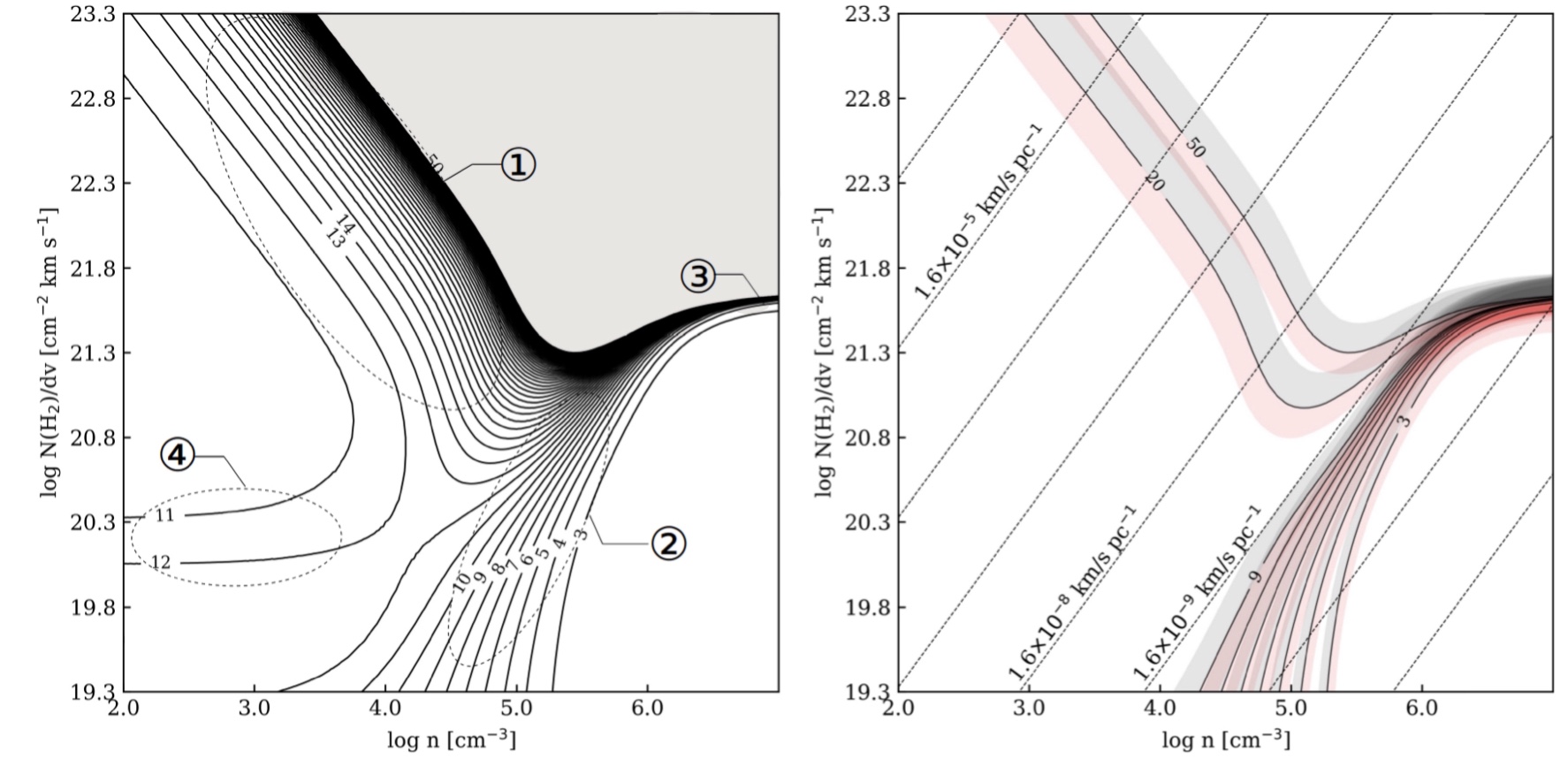}\label{fig:scatter}
\end{minipage}

\caption{\textit{Left:} The relative column density of HCN to \hcop\ (numbers on the counters), i.e., the abundance ratio, as a function of the space density (\nhtwo) and the column density (\ndvv) assuming a single component gas model with an intensity ratio of HCN-to-\hcop\ of 1.2 at 30K. The abundance of \hcop\ is fixed at [\hcop/H$_2$]~=~2$\times10^{-8}$; changing this value will shift the contour pattern along the y-axis, but the pattern shape will be the same. The four circled numbers indicate representative physical conditions that give rise to the observed intensity ratio. \textit{Right:} The relative column density of HCN to \hcop, i.e., the abundance ratio, as a function of the space density (\nhtwo) and the column density (\ndvv) assuming a single component gas model at 30K. The red shaded represents the ratio between 1.1 to 1.2 and the grey-shaded region represents 1.2 to 1.3. Dashed straight lines indicate [\hcop/H$_2$]$dv/dr$. \label{fig:1p2}
}
\end{figure*}

As {was} summarized in \S~\ref{subsec: 32}, our observations have revealed a {nearly} constant intensity ratio $\mathcal{R}$ of 1.2$\pm$0.1 across the $\sim$1~kpc nuclear bar of IC\,342 \textrm{at} a spatial resolution of 8\asec~(100~pc). Such an $\mathcal{R}$ (1.2$\pm$0.1) puts strong constraints on the physical conditions and indicates the inherent degeneracy of different physical parameters. To understand the constraints and the implication, we investigate three questions in this section. 

First, what is the relation between the parameter space of the physical conditions and a specific value of $\mathcal{R}$ (specifically, $\mathcal{R}$ of 1.2 in IC\,342, see \S~\ref{sub:r10})? {Second, 
when multiple areas in parameter space are permitted theoretically, which ones are physically realistic (\S~\ref{sub:com})?} Third, how do we reconcile conclusions from the two questions above with the $\mathcal{R}$-$L_{IR}$ relation seen in other galaxies (\S~\ref{sub:new})? {Interpreting this galactic $\mathcal{R}$-$L_{IR}$ relation requires a good understanding of how $\mathcal{R}$ is determined by the properties of the molecular gas. 
Compared to unresolved galaxy samples that established the relation,} in this paper, the constant $\mathcal{R}$ over a 1~kpc scale in IC\,342 provides constraints to the physical conditions of molecular gas from a new perspective: in other words, $\mathcal{R}$ can be relatively constant over a wide range of physical parameters. 

We discuss each of the questions in subsections below \textrm{to} conclude that $\mathcal{R}$ is a good relative abundance tracer of HCN and \hcop in IC342 and that the similar trends seen in the ratios of HCN/\hcop\ of other galaxies suggest that they may also be tracking abundance patterns there.

\subsubsection{Permitted Parameter Space for a Fixed $\mathcal{R}$} \label{sub:r10}

We investigate how a fixed value of $\mathcal{R}$ of 1.2 is realized in IC\,342 by building the radiative transfer model with different physical conditions and comparing the inferred intensity ratio, $\mathcal{R}$, to the observed values. 

First, we assume HCN and \hcop\ coexist and that the (1--0) emission lines originate from the same regions under the same temperatures and densities. The assumption is based on the similar morphology, the line profiles among HCN and \hcop\ (see \S~\ref{subsec: 32}){, and that HCN and \hcop\ have similar critical densities}. Then, we apply an escape probability formalism for a uniform sphere geometry, and use a one-component radiative transfer model with RADEX \citep{vdt07} to {explore the value of $\mathcal{R}$ over a grid of solutions in the space of the density (\nhtwo) and the column density (\ndvv)}. The optical depths and the brightness temperatures are calculated under the corresponding parameter space to determine $\mathcal{R}$ (see Appendix~\ref{appa} for details). Specifically, \nhtwo, \ndvv, and \tauu\ are all taken as constant properties within a beam.

We present in the left panel of Figure~\ref{fig:1p2} {the grid of solutions in the parameter space that produces $\mathcal{R} = 1.2$}. Axes are \nhtwo\ and \ndvv. The range of \nhtwo\ is chosen from $10^{2}$--$10^7$~\percmcu, and \ndvv\ is from $10^{21-25}$~\percmsq/50~\kms. The abundance of [\hcop/H$_2$] is fixed at 2$\times10^{-8}$ \citep[the Galactic value adopted by][]{krips08}, and the abundance of HCN is scaled relative to \hcop. Contours in the non-shadowed regions represent the column density ratios of HCN to \hcop, or equivalently, the abundance ratios that give a fixed constant $\mathcal{R}$ of 1.2. The shadowed region represents the parameter space without a solution. 

{We note that the results are insensitive to temperatures from 10 to 100 K, and show calculations for 30K, which is close to the temperature of the molecular gas suggested by \citet{mt01}.} We also find that changes in the value of the abundance of \hcop\ shift the grid of solutions along the $y$-axis (see discussions below) but do not influence our conclusions.

The non-shadowed regions in the left panel of Figure~\ref{fig:1p2} give solutions to $\mathcal{R} = 1.2$; however, not all solutions are physically realistic. We investigate the constraints on the excitation conditions or optical depth for four different \textrm{regions} marked with circled numbers on the left-hand panel of Figure~\ref{fig:1p2}. These constraints are summarized in Table~\ref{tab:region}.

\begin{table}
    \centering
    \caption{Physical conditions of the four marked regions in Figure \ref{fig:1p2}. ``sub" and ``($\sim$)therm" represent subthermalization and (near-)thermalization. See Appendix~\ref{appa} for the detailed pattern of $\tau$ and \tex\ over the (\nhtwo, \ndvv) grid.}

    \begin{tabular}{crrrr}
    \hline
    \hline
      Region  & \multicolumn{2}{c}{Optical depth} & \multicolumn{2}{c}{Excitation condition} \\
      &HCN & \hcop &HCN &\hcop \\
    \hline
        1 &\multicolumn{2}{c}{$\tau \gg 1$}  & sub & sub \\
        2 &\multicolumn{2}{c}{$\tau \sim 1$} & $\sim$therm & therm \\
        3 & $\tau$ $\gg$ 1 & $\tau$ $\sim$ 1.8  & therm & therm \\
        4 &  $\tau$ $\gg$ 1 & $\tau \gtrsim$ 1 & sub & sub\\
         \hline
    \end{tabular}
    \label{tab:region}
\end{table}

In region 1, both lines are subthermalized and collisions control the level populations. Both lines are very optically thick, and the intensity ratio {is determined by} the ratio of the excitation temperature \tex. Region 2, in contrast, controls $\mathcal{R}$ through the relative optical depth together with \tex. In region 2, both HCN and \hcop\ have intermediate optical depth, and \hcop\ emission is thermalized. {In region 3, the particle density is very high, and both lines are thermalized and \tex\ of HCN and \hcop\ are equal. Therefore, the ratio of $1-e^{\tau}$ is a constant. Region 3 thus requires a constant $\tau$(\hcop) for which \mbox{$1-e^{-\tau}$ = 1/1.2}, because HCN emission is optically thick.} Such a tight constraint rules out region 3 as a realistic solution. {Finally, contours in region 4 are parallel to the horizontal axis. While both HCN and \hcop\ are subthermalized in this region due to the low value of \nhtwo,  both $\tau$(\hcop) and the \tex\ ratio are independent of \nhtwo\ in this region. Similarly to region 3, we consider the constant $\tau$ in region 4 to be unrealistic.}

Since both regions 1 and 2 are the permitted parameter spaces for a fixed $\mathcal{R}$ with a value of 1.2, we further compare the two regions in \S~\ref{sub:com} to see whether they \textrm{can} apply to the whole bar region of IC\,342.

\subsubsection{Thermal vs. Subthermal in the Bar of IC\,342}\label{sub:com}

{Our RADEX modeling results show that either region 1 or region 2 are permitted solutions of parameter space for the constant $\mathcal{R}$ across the IC\,342 bar. To further determine whether both regions are equally likely or that one region is more likely than the other, we compare the theoretical predictions with the observations, including the column densities, the optical depths, and the abundance ratios derived from current and/or past studies as direct evidence. As we will conclude below, region 2 (thermal excitation) serves this requirement better than region 1 (subthermal excitation) does. We summarize the key criteria in Table~\ref{tab:412}.

\begin{table*}
    \centering
    \tabletypesize{\scriptsize}
    \caption{Key criteria for identifying the physically more feasible parameter space in the bar region of IC\,342.}

    \begin{tabular}{llc}
    \hline
    \hline
     Comparison with ... &  Criteria & Better Region\\
     \hline
     Models & Range of [\hcop/H$_2$]$dv/dr$ & Region 2\\
     & Allowed uncertainties for a specific $\mathcal{R}$ & Region 2\\
     Observations & Column density measured by \ceio(1--0) & Region 2 \\
     & \tauu(HCN) estimated by comparing single-dish/interferometer maps & Region 2\\
     &\tauu(HCN) derived through H$^{12}$CN(1--0)/H$^{13}$CN(1--0) & {N/A}\\
     & The abundance ratio of HCN-to-\hcop\ & N/A \\
     \hline
    \end{tabular}
    \label{tab:412}
\end{table*}

\paragraph{Theoretical constraints from the RADEX modeling} A constant $\mathcal{R}$ in the grid of (\nhtwo, \ndvv) is produced along the abundance ratio contours (Figure~\ref{fig:1p2}). The orientation of the contours reflects certain relationships of \nhtwo\ and \ndvv. For example, in region 1, the contour pattern indicates that $n_{\textrm{H}_2}\cdot N\sim$~constant. In region 2, the contour lines resemble more of the condition that $dv/dr$ is constant, because \ndvv$\cdot dv/dr \sim N_{\textrm{H}_2}/dr \sim$ \nhtwo. This point is illustrated as a group of parallel straight lines with a slope of one in the right panel of Figure~\ref{fig:1p2}. The intercept of each line there corresponds to a different [\hcop/H$_2$]$dv/dr$. As is shown in Figure~\ref{fig:1p2}, while [\hcop/H$_2$]$dv/dr$ only spans one order of magnitude from 1.6$\times10^{-10}$ to 1.6$\times10^{-9}$~\kms~pc$^{-1}$ in region 2, in region 1 the range is as high as four orders of magnitude. Therefore, the \nhtwo-\ndvv\ relation in region 2 is more realistic for a fixed $\mathcal{R}$. {We note that we justify the feasible parameter space with the product of [\hcop/H$_2$] and $dv/dr$, rather than [\hcop/H$_2$] or $dv/dr$ separately because we do not have much knowledge about each of them.} The ``indirect" evidence of a constant HCN(1--0)-to-\hcop(1--0) ratio across the IC\,342 bar fits the movement of points along the abundance ratio contours of region 2 better. If points on the (\nhtwo, \ndvv) space move along the constant $dv/dr$ direction in the grid, region 1 has a smaller tolerance on the uncertainties to the solutions of a specific $\mathcal{R}$. We illustrate this in the right panel of Figure~\ref{fig:1p2}, in which we only keep contour lines in regions 1 and 2 for clarity. Specifically, the red- and grey-shadowed regions represent $\mathcal{R}$ from 1.1 to 1.2, and from 1.2 to 1.3, a range that represents the uncertainty level we have observed in IC\,342. While the wider shaded region in region 1 seems to allow for a larger permitted parameter space under $\mathcal{R}$ from 1.1 to 1.3, the permitted range along the constant $dv/dr$ is actually much smaller than that of region 2. 

\paragraph{Observational constraints} We also consider observational constraints on the column densities, the optical depths, and the abundance ratio to discuss which region on the grid of (\nhtwo, \ndvv) is the preferred parameter space for the bar of IC\,342. 

We \textrm{make the assumption that \ceio\ and HCN (or \hcop) trace} the same gas to compare column densities and decide which excitation region better represents the conditions in IC~342.  As we have shown in \S~\ref{subsec: 32}, the corresponding $ N_{\textrm{H}_2}$ (or \ndvv) is $\sim 10^{22}$~\percmsq~(or 10$^{20.3}$~\percmsq~\kms). This value agrees better with the column density referred to by region 2.

The comparison between the single-dish GBT/\textit{Argus} image and the interferometric images shows that HCN has an intermediate optical depth. {In observations by interferometers PdBI in \citet{downes92} and \citet{sch08}, several (3--5) individual giant molecular clouds (GMCs) on scales of ~80 pc were found encircling a $\sim$100~pc hole at the center of IC\,342. If material on the near side of the nucleus had high optical depth, the GMCs revealed by the interferometers would then be behind an opaque screen and would be invisible{, either spatially or in the velocity space}. An alternative explanation is that the foreground screen consists of a collection of high opacity but clumpy HCN cores {not yet resolved by the interferometer (e.g., $\sim$1 pc scales}, and the gaps between the cores are much smaller than the GBT beam.}


{Results from past observations of isotopologues of HCN in IC\,342 can be consistent with either region 1 or region 2 ~\citep{downes92, wilson99, schulz01, sch08}. \citet{schulz01} measured H$^{12}$CN(1--0)/H$^{13}$CN(1--0) over GMCs in the center of IC\,342 and found a value of  $\sim40$. This value is larger than one, and even larger than the local [$^{12}$C/$^{13}$C] ratio of $\sim24$ \citep{halfen17}. These measurements indicate the partial coverage of H$^{13}$CN relative to H$^{12}$CN must be smaller than one. A fractional coverage of 0.1 would result in a high optical depth of H$^{12}$CN, and a fractional coverage of 0.4 would give H$^{12}$CN an intermediate optical depth ($\tau\sim1$). }





\textrm{A known abundance ratio of HCN to \hcop\ could distinguish between regions 1 and 2 if the range of its values could be constrained to {order of unity (region 2) or order of ten (region 1)}.} Theoretical models can predict the abundance ratio of HCN to \hcop. For example, shocked molecular clouds allow possible endothermic reactions to occur and gas-phase ion-molecule reactions to proceed in the compressed layers, so the abundance of \hcop\ could be suppressed by at most two orders of magnitudes \citep[see Figure~1 in][]{ig78}. As a comparison, models of PDRs mostly yield [HCN/\hcop]$>$1 \citep{papadopoulos07}. Since bar-induced shocks and PDR regions produced by the central starburst \citep{meier05} coexist in the nucleus of IC\,342, we cannot distinguish region~1 and 2 through the predictions on the abundance ratio.

\subsubsection{Thermal vs. Subthermal in the General Case} \label{sub:new}

We \textrm{described} in \S~\ref{sub:r10} and Table~\ref{tab:412} that we favor region 2 {more than} region 1 in the nuclear bar region of IC\,342. The most common interpretation for  $\mathcal{R}$ in the literature, however, is subthermal excitation together with large optical depth, or the physical conditions indicated by region 1. As we will describe in detail below, this interpretation is often used for determining the $\mathcal{R}$-$L_{IR}$ or the $\mathcal{R}$-AGN/SB relationships, whereas these relationships are typically inferred from unresolved or low spatial resolution observations \citep[e.g.,][]{krips08, jimnez17}.

The degeneracy in solution between regions 1 and 2 that we have seen in IC\,342 is unlikely to be unique to this galaxy (Table~\ref{tab:412}). Therefore, we investigate (1) whether region 1 is the only possible solution to single-beam observations on samples of external galaxies, and (2) whether region 2 serves as a proper solution if external galaxies are spatially resolved and still have a constant $\mathcal{R}$.

We first revisit how region 1 was derived from unresolved or low spatial resolution observations. Take NGC\,253 as an example: it has spatially resolved and indistinguishable HCN(1--0) and \hcop(1--0) maps~\citep{knudsen07}, and a constant $\mathcal{R}$ of 1.4 is measured for the two maps. \citet{knudsen07} compare the (1-0) maps of NGC 253 to the single beam (3-2) data and conclude that both HCN(1--0) and \hcop(1--0) are optically thick and are subthermally excited (\nhtwo\ = 10$^{5.2}$~\percmcu\ at 50~K).

We can derive the same conclusion for IC\,342 using the same analysis method. For IC\,342, the James Clerk Maxwell Telescope (JCMT) observed $J$=4--3 transitions of HCN and \hcop~\citep{tan18}. We convolved the $J$=1--0 image to the 14\asec\ beam resolution of HCN(4--3), and found that the line intensity ratio of \mbox{(1--0)/(4--3)} (hereafter referred as $r_{41}$) is $\sim$3--10 for HCN, and $\sim$2.5--6 for \hcop, respectively. Using the same parameters in RADEX as \S~\ref{sub:r10} over the one-component model, the range of $r_{41}$ corresponds to \nhtwo\ from 10$^{4.5}$--10$^{5.5}$~\percmcu\ (Figure~\ref{fig:highj}) if HCN(1--0) is optically thick (\tauu$\gg$1). This result coincides with region~1  (see Figure~\ref{fig:1p2}) and seems to be consistent with the condition that \citet{knudsen07} have suggested.

\begin{figure}
    \centering
    \includegraphics[width=\linewidth]{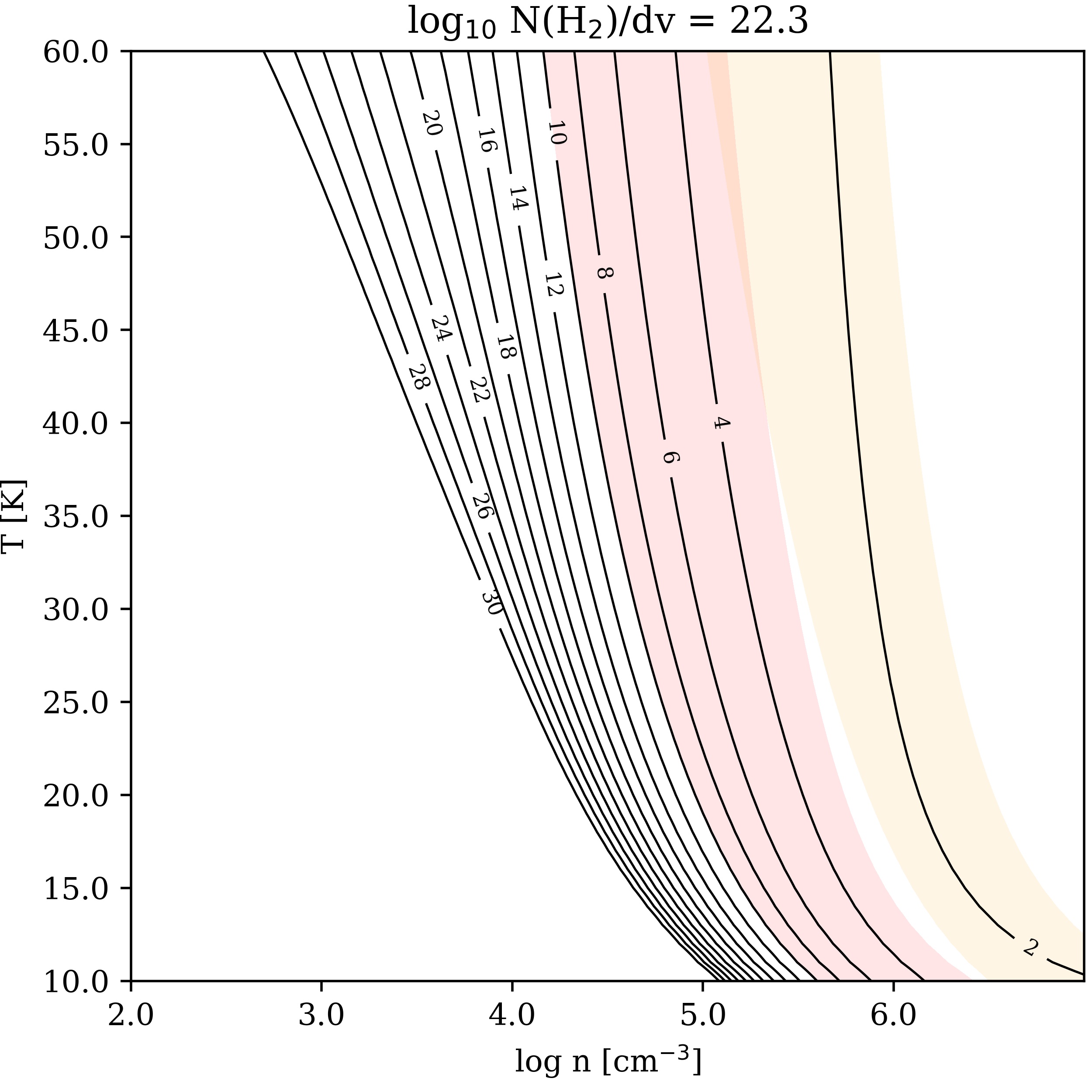}
    \caption{Contours showing $r_{41}$ of HCN over the grid ($n$, $T$) for log$_{10} N(H_2)/dv$ = 22.3. The red region shows where  $r_{41}$ is from 3 to 10. The orange area shows the solution for $r_{41}$ from 3 to 10 when log$_{10} N(H_2)/dv$ = 21.3, which has a 10 times lower column density.}
    \label{fig:highj}
\end{figure}

\begin{figure*}[!t]
\begin{minipage}[t]{\textwidth}
\includegraphics[width=\textwidth]{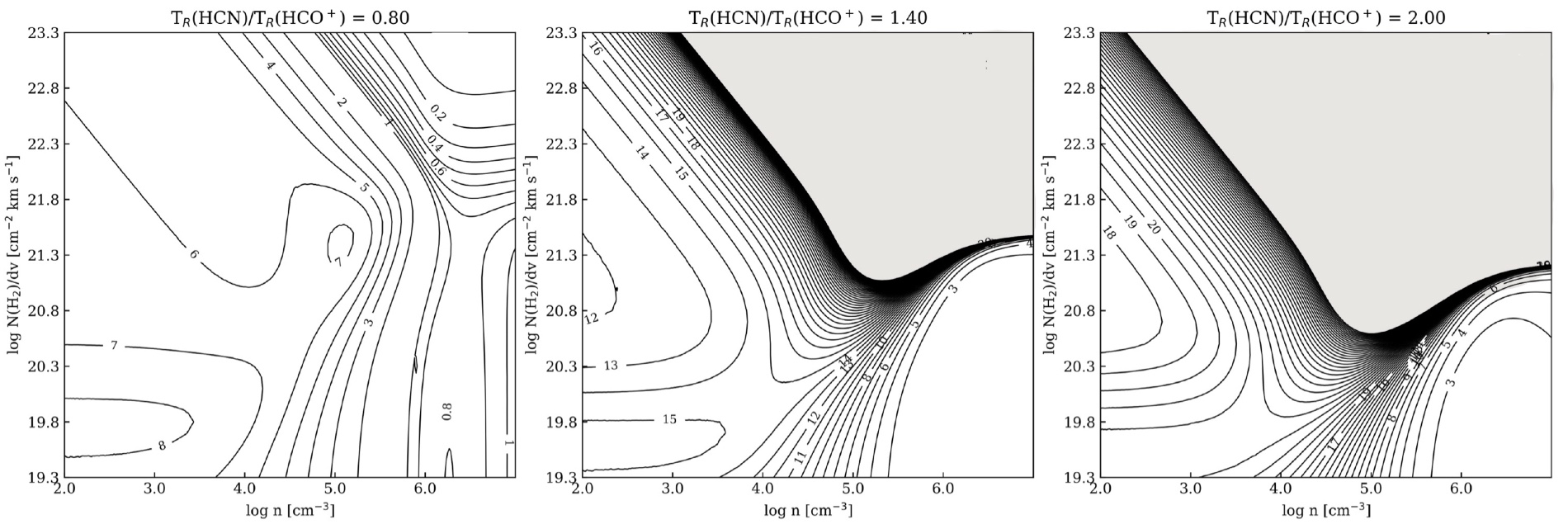}
\end{minipage}

\caption{Assuming a one-component model, the available parameter space (the non-shadowed regions) corresponding to the intensity ratio of HCN-to-\hcop\ of 0.8, 1.4, and 2 under 30 K. Contours represent the relative column density of HCN to \hcop, i.e. the abundance ratio. \label{fig:four-grid}}
\end{figure*}

However, such a combination of the column density and the particle density is not the unique solution that comes from single-beam observations. For example, a combination of a column density that is 10 times lower, an intermediate optical depth close to one, and a high \nhtwo\ at which HCN(1--0) is thermalized also produces $r_{41}$ in range of $\sim$3--10 (see the orange area in Figure~\ref{fig:highj}).

We, therefore, conclude that region 1 may not be the only possible solution to single-beam observations of external galaxy samples. This is because, in the analysis above, single-component modeling of emission from states with very different excitation conditions is likely an oversimplification. First, $r_{41}$ is {insensitive} to either temperature or [HCN/H$_2$]$/(dv/dr)$, but is very sensitive to \nhtwo~(see Figure~\ref{fig:highj}). A small variation in the value of the line intensity ratio will change the result of \nhtwo\ significantly and will bring a large uncertainty on the derived \nhtwo. Second, the one-component model used in analyzing smoothed intensity ratios in a large beam does not represent the actual physical conditions, because high-$J$ lines require much higher critical densities for \textrm{thermalization}. While $r_{41}$ is tracing more of the bright $J$=4--3 lines, it probes more of the fraction of high excitation materials within the beam, or equivalently, the {mean} particle density. 

{We next investigate whether region 2 serves as a proper solution to situations where external galaxies are spatially resolved and still have a constant $\mathcal{R}$.} 
We examine solutions of the (\nhtwo, \ndvv) parameter space when $\mathcal{R}$ is between 0.8 to 2, a range that is typical in a sample of galaxies when the $\mathcal{R}$-$L_{IR}$ or $\mathcal{R}$-AGN/SB relations are discussed \citep[e.g.,][]{krips10}. Figure~\ref{fig:four-grid} presents the solutions to $\mathcal{R}$ = 0.8, 1.4, and 2. For $\mathcal{R}$ = 0.8 (or that $\mathcal{R}$ is smaller than one), all (\nhtwo, \ndvv) pairs have a solution for the abundance ratio. This is because \hcop\ has a smaller critical density and is thermalized more easily than HCN. If $\mathcal{R}$ is larger than one, the solution in the (\nhtwo, \ndvv) grid is similar to that of $\mathcal{R}$ = 1.2 (see Figure~\ref{fig:1p2}). Specifically, as $\mathcal{R}$ increases, the upper limit of the allowed solution for \ndvv\, at \nhtwo~$< 10^5$~\percmcu\ decreases, and the corresponding abundance ratio for (\nhtwo, \ndvv) at \nhtwo~$< 10^5$~\percmcu\ increases (see Figure~\ref{fig:four-grid}). Such a behavior explains why the value of $\mathcal{R}$ as large as, for example, 5, is not seen in external galaxies. For such a high value of $\mathcal{R}$, there is no solution in the (\nhtwo, \ndvv) grid at \nhtwo~$< 10^5$~\percmcu. At \nhtwo~$> 10^5$~\percmcu, the (\nhtwo, \ndvv) solution requires an abundance ratio of $\sim$50 that is unlikely. {This is consistent with conclusions from \citet{yamada07} or \citet{gg08} that very high HCN/\hcop\ intensity ratios require extreme abundance ratios.}

Figure~\ref{fig:four-grid} shows that $\mathcal{R}$ being smaller than one is not a peculiar result and does not put strong constraints on physical conditions. In contrast, $\mathcal{R}$ with a value larger than one can be used to constrain physical conditions. {Therefore, for external galaxies that are spatially resolved and have a constant $\mathcal{R}$ distribution with values larger than one, we can conclude that region 2 is still the preferable parameter space because we require a parameter space that is {insensitive} to the change of physical conditions across a large physical scale, which refers to the nuclear bar region regarding IC\,342 
(see \S~\ref{sub:com}). Furthermore, we emphasize that the corresponding abundance ratio in region 2 changes under different $\mathcal{R}$. This implies that $\mathcal{R}$, or the HCN-to-\hcop\ intensity ratio, is more sensitive to small changes in the abundance ratio than changes in temperature or particle density. }



The idea that the HCN-to-\hcop\ intensity ratio traces abundance is related to using $\mathcal{R}$ as a potential diagnostic tool to distinguish AGN and starbursts in galactic nuclei. It has been argued that the abundance ratio of HCN to \hcop\ is sensitive to the environment which hosts an AGN or a starburst {\citep[e.g.,][]{gg06, ima06, ima07}}. First, HCN can be enhanced chemically either by FUV radiation from young massive star-forming regions or through strong X-ray radiation from an AGN. AGN are therefore likely to have a higher HCN abundance since X-ray radiation penetrates more deeply into gas clouds than UV radiation. Second, evolved starbursts tend to have a lower HCN-to-\hcop\ abundance ratio due to the ionization effects from cosmic rays from supernovae. The ionization effects potentially increase the \hcop\ abundance and decrease the HCN abundance \citep[e.g.,][]{behrens22, krips08, meij11}.

In conclusion, (1) region 1 does not uniquely account for multi-line constraints from single-dish observations, and (2) samples of external galaxies in the $\mathcal{R}$-$L_{IR}$ or $\mathcal{R}$-AGN/SB relation, if they are mapped with finer resolution, might still have a constant $\mathcal{R}$ across HCN/\hcop\ emitting region. And if $\mathcal{R}$ is larger than one, region~2 is the preferable parameter space. In this case, $\mathcal{R}$ is insensitive to the exact condition of the two species and is telling more of the relative abundance of the two species. This in turn relates to the existence of AGN or starbursts. 

\begin{figure*}
    \centering
\includegraphics[width=0.8\textwidth]{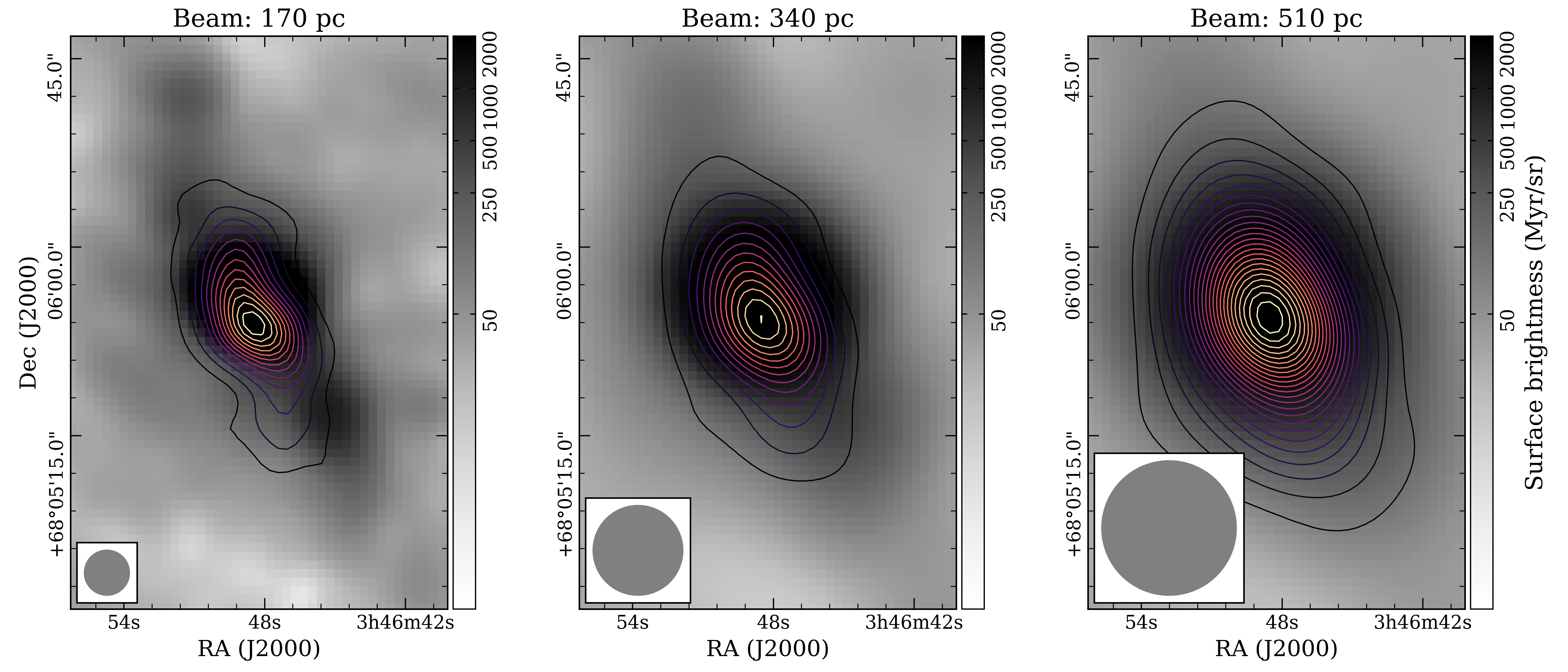}
    \caption{Comparion between the HCN moment 0 maps and the \textit{Herschel} PACS 70~\um\ image~\citep{kennicutt11} under a beam resolution of 10.7\asec, 21.4\asec, and 32.1\asec~(170, 340, and 510~pc). Contours start from 3$\sigma$ and are in steps of 3$\sigma$. The correspondent 1$\sigma$ levels are 0.5, 0.35, 0.13 K~\kms.}
    \label{fig:hcn-70-map}
\end{figure*}

\begin{figure*}[!t]
\begin{minipage}[t]{\textwidth}
\centering
\includegraphics[width=\textwidth]{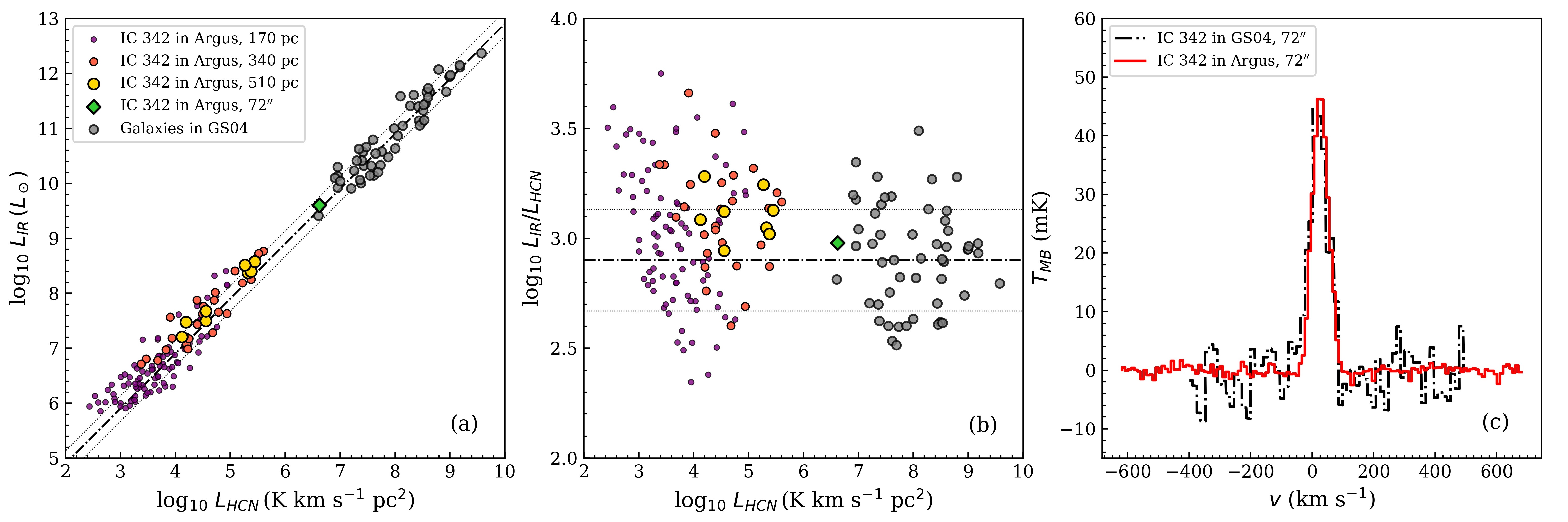}
\end{minipage}

\caption{(a) and (b): $L_\textrm{IR}$-$L_\textrm{HCN}$ relations from individual sampling points on the IC\,342 \textit{Argus} image on a scale of 170 (\textit{purple}), 340 (\textit{orange}), and 510 pc (\textit{yellow}). Each individual point in the figure shows the luminosities of individually sampled apertures. The central 72\asec\ beam in the \textit{Argus} image ($\sim$1150~pc, \textit{green}); $L_\textrm{HCN}$ of which is converted from the red spectrum on Figure~\ref{fig:ir-hcn-all}(c) is shown to compare with the beam size of measurements in \citet{gs04}  (\textit{grey}). The dashed-dotted and dotted lines represent the best fitted $L_\textrm{IR}$-$L_\textrm{HCN}$ relation and the corresponding 1$\sigma$ level reported in \citet{gs04}. (c) Spectra observed by \textit{Argus} and by \citet{gs04} in \tmb\ scale from a 72\asec\ beam at the center of IC\,342 at RA = 3:46:48.30795, DEC = +68:05:48.8251 (J2000). \label{fig:ir-hcn-all}}
\end{figure*}

\subsection{Breakdown of the HCN-IR Correlation at High Spatial Resolution}

On a galactic scale, a scale-dependent scatter was previously observed with CO in the relation between the molecular gas surface density and the SFR \citep[e.g.,][]{schruba10, momose13, kruijssen14, kreckel18, pan22}. Such a relationship breaks down on smaller {spatial} scales due to incomplete sampling of star-forming regions {by CO emission. \textit{Argus} on the GBT, on the other hand, with its relatively high (6--8\asec) spatial resolution and excellent surface brightness sensitivity, provides a unique opportunity to inspect a similar breakdown in the \mbox{$L_\textrm{IR}$-$L_\textrm{HCN}$} relationship in nearby HCN-bright galaxies as the spatial resolution decreases (Figure~\ref{fig:hcn-70-map}). In earlier studies with the IRAM 30~m telescope, the EMPIRE survey~\citep{jimnez17}, at 30\asec\ resolution, reveals that the ratios of HCN/CO and HCN/IR show correlations with radius, galaxy surface mass density, molecular gas surface density, and gas pressure.}


We present in Figure~\ref{fig:ir-hcn-all}a and \ref{fig:ir-hcn-all}b the \mbox{$L_\textrm{IR}$-$L_\textrm{HCN}$} and the \mbox{$L_\textrm{IR}$/$L_\textrm{HCN}$-$L_\textrm{HCN}$} relation on scales of 170~pc ($\sim$10\asec), 340~pc ($\sim$20\asec), and 510 pc ($\sim$30\asec) obtained by convolving our \textit{Argus} HCN image to different beam sizes. The HCN luminosity, $L_\textrm{HCN}$, was derived from the integrated intensities following \citet{gs04}. We confirm that the measurements of the line intensities from the two studies are consistent by presenting in Figure~\ref{fig:ir-hcn-all}c the HCN spectra observed from a 72\asec\ beam at the same position, although we note that there is a blue-shifted wing in the \citet{gs04} spectrum that we do not detect. {The total infrared luminosity, compared to IRAS data used in \citet{gs04}, was calibrated from the brightness in the \textit{Herschel} 70~$\mu$m band \citep{kennicutt11}, adopting calibration coefficients from \citet{gala13}. The converted $L_\textrm{IR}$ is consistent with that in \citet{gs04} derived from IRAS data (Figure~\ref{fig:ir-hcn-all}a and \ref{fig:ir-hcn-all}b).} All the data points on (sub-galactic) scales from 170 to 510~pc fall on the \citet{gs04} relation with a slope of one (Figure~\ref{fig:ir-hcn-all}a and \ref{fig:ir-hcn-all}b). 

Table~\ref{tab:scatter} summarizes the variation of the scatter, i.e., the standard deviation of log$_{10}(L_\textrm{IR}/L_\textrm{HCN})$, of \textit{Argus} data to the $L_\textrm{IR}$-$L_\textrm{HCN}$ relation of which the power is one under different sampling sizes. The scatter drops as the spatial scale increases. The scatter for the smallest (170~pc) scale of the corresponding $L_\textrm{IR}$-$L_\textrm{HCN}$ relation is only \mbox{1.3 times} the scatter for beam-averaged external galaxies, which is comparable to the scatter at 340~pc in IC\,342. {As Figure~\ref{fig:12co} shows, \textit{Argus} resolves the angular offset between the IR emission and molecular gas emission with a 100~pc resolution. The scatter at a 340~pc scale therefore intrinsically reflects this angular offset. } 

\begin{table}[!t]
\caption{The average and the scatter in $\textrm{log}_{10}(L_\textrm{IR}/L_\textrm{HCN})$ in different spatial scales.}
\centering

\begin{tabular}{ccc}
\hline
\hline
Regions & $\langle \textrm{log}_{10}(L_\textrm{IR}/L_\textrm{HCN}) \rangle$ & $\sigma( \textrm{log}_{10}(L_\textrm{IR}/L_\textrm{HCN}))$  \\
\hline
170 pc& 3.02 & 0.31\\
340 pc& 3.10 & 0.23\\
510 pc& 3.11 & 0.10\\
\hline
GS04\tablenotemark{a} & 2.92 & 0.23 \\
\hline
    \end{tabular}
    \label{tab:scatter}
    \tablenotetext{a}{Values adopted from \citet{gs04}.}
\end{table}

The scale-scatter breakdown observed in CO is interpreted as an incomplete sampling of star-forming regions. Specifically, \citet{schruba10} propose that the breakdown reflects a temporal evolution process of gas depletion in star formation revealed by different spatial scales. However, we may not conclude that the breakdown in the $L_\textrm{IR}$-$L_\textrm{HCN}$ relation of IC\,342 is also due to the gas depletion in star formation because the shift between $L_\textrm{IR}$ and $L_\textrm{HCN}$ is from the bar rotation, as the evolution in IC\,342 is possibly influenced by the passing density wave~\citep{turner92}. We can only argue that the HCN(1--0) emission is more strongly correlated with star formation when averaged over large regions of IC\,342 than measured at high resolution. The high-resolution observations suggest that the relationship begins to break down on smaller 340~pc regions. {As a reference, we note that \citet{murphy15} has reported a spatial offset between HCN (and \hcop) with the continuum tracing star-formation on a physical scale of $\sim$130~pc in NGC~3627.} 

We conclude that HCN(1--0) emission can serve as a good star formation tracer if a certain spatial sampling criterion is satisfied, although this may not fully reflect a dense gas-to-star formation relation. More samples with spatially resolved HCN emission from the DEGAS survey should provide a further understanding of the criteria that sustain the tight global $L_\textrm{IR}$-$L_\textrm{HCN}$ relation.

\section{Summary}\label{sec:summary}

We have observed ground state transitions of \co, \thco, \ceio, HCN, and \hcop\ from the external galaxy IC\,342 with the 16-pixel spectroscopic focal plane \textit{Argus} array on the 100-m GBT. These data provide single-dish measurements at high spatial resolution (6--10\asec, $\sim$100~pc). The gaseous nuclear bar is mapped in all five transitions and \co(1--0) observations reveal the two inner spiral arms. The morphology of molecular gas traced by \co(1--0) and \thco(1--0) differs from that of HCN(1--0) and \hcop(1--0), indicating different sensitivity of these molecular gas tracers to different physical conditions. {\thco(1--0) traces the surface layer of molecular clouds as \co(1--0) does due to radiative trapping.} HCN emission correlates well with infrared emission tracing recent star formation as long as a sufficiently large sampling scale{ (e.g. $>$340~pc for IC\,342)} is satisfied.

Our \textit{Argus} observations show that the {intensity of} spatially resolved HCN(1--0) and \hcop(1--0) emission have a remarkably tight correlation, $\mathcal{R}$ of $ 1.2\pm0.1$, with independent measurements at 100~pc resolution over the entire 1~kpc bar. This suggests that the line ratio of HCN(1--0) to \hcop(1--0) is {insensitive to} local varying physical conditions across the center of IC342. We use RADEX with a one-component model to explore the permitted parameter space over this constant line ratio and identify the preferred region to realize the insensitivity of the line ratio to the environments. Our investigation also searches for available parameter space for HCN(1--0)/\hcop(1--0) ratios observed among other external galaxies. For IC\,342, we conclude that the HCN(1--0) and \hcop(1--0) emission from the 1~kpc gaseous bar of IC\,342 likely have intermediate optical depth and \nhtwo\ between 10$^{4.5}$--10$^{6}$~\percmcu, where \hcop(1--0) is thermalized and HCN(1--0) is close to thermalization. This result is compatible with results derived from analyses with higher-$J$ transitions, {although it }also indicates that the HCN(1--0) and \hcop(1--0) emission from external galaxies may have physical conditions other than {large optical depth and specific particle densities that cause subthermalization.} {The insensitivity of the HCN(1--0)-to-\hcop(1--0) intensity ratio across the center of IC\,342 indicates that it is more sensitive to the relative abundance ratio rather than particle densities.}

The high spatial resolution of \textit{Argus} also provides a unique opportunity to inspect the breakdown between the sampling scale and the degree of scatter of the $L_\textrm{IR}$-$L_\textrm{HCN}$ relation in the nearest HCN-bright galaxies. We find that the scatter of the $L_\textrm{IR}$-$L_\textrm{HCN}$ relationship decreases as the spatial scale increases from 10\asec\ to 30\asec\ (170–510 pc) and is comparable to the scatter of the global relation at the scale of 340 pc.
 
We would like to thank Dr. Stuart Vogel for useful and constructive discussions, Dr. Karen O'Neil (GBT) for the approval of the DDT request, and the operator teams for their help in the observations. We appreciate the constructive suggestions from the anonymous referee to improve the quality of the manuscript. This work was conducted as part of the ``Dense Extragalactic GBT+\textit{Argus} Survey" (DEGAS) collaboration. We thank the \textit{Argus} instrument team from Stanford University, Caltech, JPL, University of Maryland, University of Miami, and the Green Bank Observatory for their efforts on the instrument and software that have made this work possible. The \textit{Argus} instrument construction was funded by the National Science Foundation (NSF) ATI-1207825. The authors acknowledge funding from the award NSF AST-1615647 to the University of Maryland, and NSF AST-1616088 to the University of Miami. The National Radio Astronomy Observatory is a facility of the National Science Foundation operated under cooperative agreement by Associated Universities, Inc.

\software{Astropy \citep{astropy13, astropy18}, APLpy (\url{http://aplpy.github.com}), degas (\url{https://github.com/GBTSpectroscopy/degas}), GBTIDL (\url{https://gbtidl.nrao.edu/}), gbtpipe (\url{https://github.com/GBTSpectroscopy/gbtpipe}), NumPy \citep{harris20}, SciPy \citep{virtanen20}, Spectral Cube (\url{https://github.com/radio-astro-tools/spectral-cube}.)}

\appendix
\section{Solutions for a specific line intensity ratio}\label{appa}

\begin{figure*}[!ht]
\begin{minipage}[t]{\textwidth}
\centering
\includegraphics[width=1\textwidth]{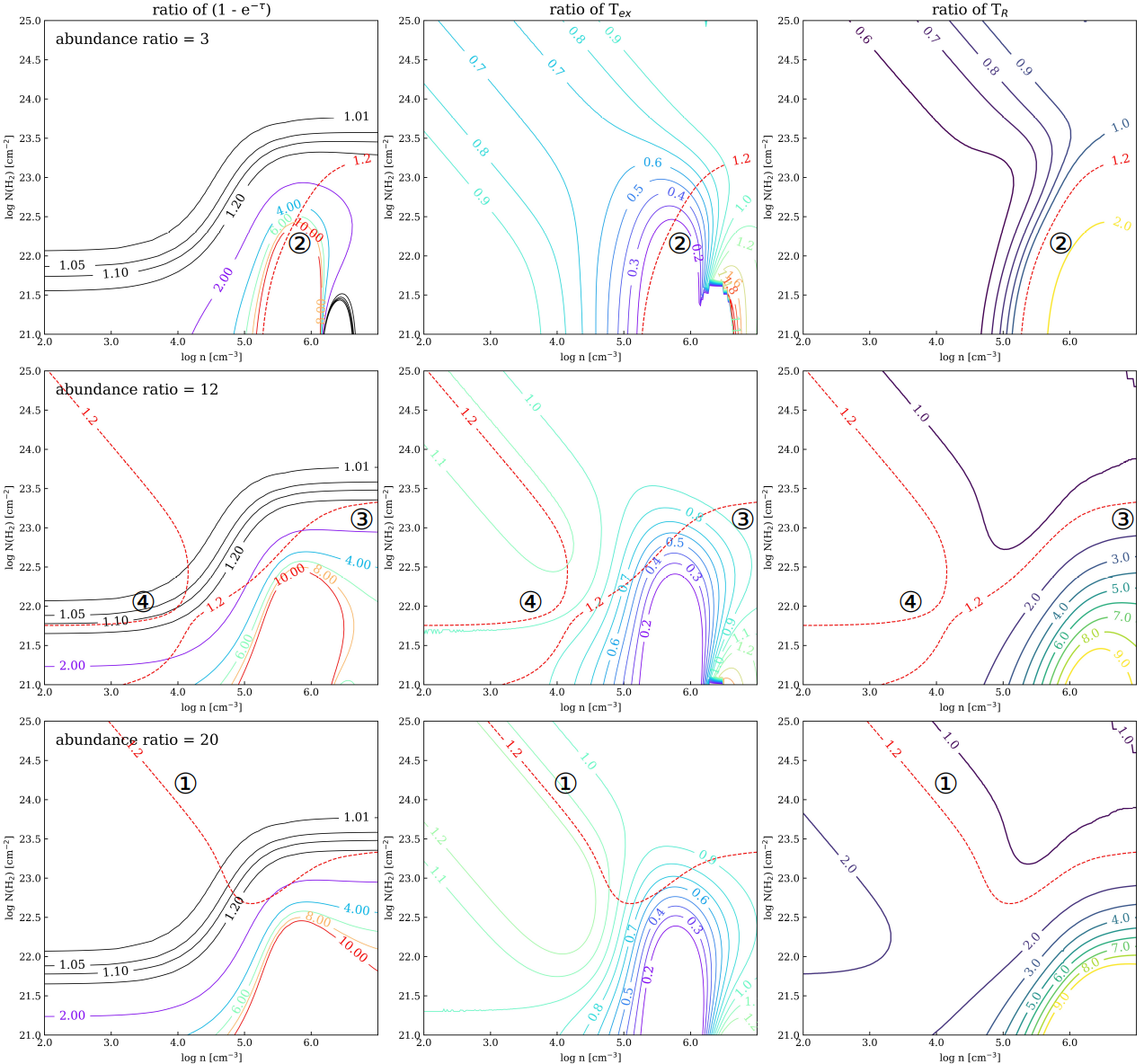}
\end{minipage}

\caption{The optical depth effects on the emission, and the excitation condition of HCN and \hcop\ in a 30 K environment. Each row presents a representative relative column density of HCN to \hcop\ of 3, 12, and 20. The first column shows the ratio of ($1-e^{-\tau}$) of HCN and \hcop. The second column shows the ratio of the excitation temperature of both molecules, HCN to \hcop. The third shows the ratio of the radiation temperature, which is $T_{ex}(1-e^{-\tau})$. The red dashed line in each panel indicates where the line intensity ratio is 1.2, assuming that HCN and \hcop\ have the same line width. Labels of regions are consistent with those in Figure \ref{fig:four-grid}. \label{fig:abr-detail}}
\end{figure*}

We plot in Figure \ref{fig:abr-detail} the ratio of ($1-e^{-\tau}$), and the ratio of \tex\ between HCN(1--0) and \hcop(1--0) in the first and the second column. The optical depth effects and the excitation condition are represented, sequentially. Plots in the third column show the joint result of $\tau$ and \tex, as $T_\textrm{R}$~=~$(1-e^{-\tau}) J(T_\textrm{ex})$. $J(T_\textrm{ex})$ is the Rayleigh-Jeans corrected excitation temperature. 

We overlay the contour representing a $T_\textrm{R}$ ratio of 1.2 on the first two columns to see the corresponding $\tau$ and \tex. The four representative regions discussed in Section~\ref{subsec:ratio} are labeled in circled numbers: (1) region 1 is contributed by very large optical depths, as it is located where the ratio of ($1-e^{-\tau}$) of HCN and \hcop\ is close to 1, and therefore the ratio of \tr\ equal to 1.2 is because the ratio of \tex\ is 1.2, (2) region 2 spans across where HCN is less thermalized than \hcop, and its brightness is contributed by larger ($1-e^{-\tau}$), (3) region 3 is where both HCN and \hcop\ are thermalized, and the ratio of ($1-e^{-\tau}$) is fixed to 1.2 regardless of the abundance ratio, and (4) region 4 follows the contours when the ratio of \tex\ is slightly larger than 1 while both lines are subthermalized, and ($1-e^{-\tau}$) is constant.


\begin{thebibliography}{}









\bibitem[Astropy Collaboration et al.(2013)]{astropy13} Astropy Collaboration, Robitaille, T.~P., Tollerud, E.~J., et al.\ 2013, \aap, 558, A33. doi:10.1051/0004-6361/201322068

\bibitem[Astropy Collaboration et al.(2018)]{astropy18} Astropy Collaboration, Price-Whelan, A.~M., Sip{\H{o}}cz, B.~M., et al.\ 2018, \aj, 156, 123. doi:10.3847/1538-3881/aabc4f

\bibitem[Behrens et al.(2022)]{behrens22} Behrens, E., Mangum, J.~G., Holdship, J., et al.\ 2022, \apj, 939, 119. doi:10.3847/1538-4357/ac91ce

\bibitem[Bigiel et al.(2008)]{bigiel08} Bigiel, F., Leroy, A., Walter, F., et al.\ 2008, \aj, 136, 6

\bibitem[Bigiel et al.(2016)]{bigiel16} Bigiel, F., Leroy, A.~K., Jim{\'e}nez-Donaire, M.~J., et al.\ 2016, \apjl, 822, L26. doi:10.3847/2041-8205/822/2/L26

\bibitem[Bolatto et al.(2013)]{bolatto13} Bolatto, A.~D., Wolfire, M., \& Leroy, A.~K.\ 2013, \araa, 51, 207. doi:10.1146/annurev-astro-082812-140944

\bibitem[Crosthwaite et al.(2000)]{crosthwaite00} Crosthwaite, L.~P., Turner, J.~L., \& Ho, P.~T.~P.\ 2000, \aj, 119, 1720. doi:10.1086/301302


\bibitem[Crosthwaite et al.(2001)]{crosthwaite01} Crosthwaite, L.~P., Turner, J.~L., Hurt, R.~L., et al.\ 2001, \aj, 122, 797

\bibitem[Dahmen et al.(1998)]{dahmen98} Dahmen, G., Huttemeister, S., Wilson, T.~L., et al.\ 1998, \aap, 331, 959



\bibitem[Downes et al.(1992)]{downes92} Downes, D., Radford, S.~J.~E., Guilloteau, S., et al.\ 1992, \aap, 262, 424

\bibitem[Eckart et al.(1990)]{eckart90} Eckart, A., Downes, D., Genzel, R., et al.\ 1990, \apj, 348, 434

\bibitem[Egusa et al.(2011)]{egusa11} Egusa, F., Koda, J., \& Scoville, N.\ 2011, \apj, 726, 85

\bibitem[Elmegreen(2002)]{elmegreen02} Elmegreen, B.~G.\ 2002, \apj, 577, 206. doi:10.1086/342177

\bibitem[Evans et al.(2001)]{evans01} Evans, N.~J., Rawlings, J.~M.~C., Shirley, Y.~L., et al.\ 2001, \apj, 557, 193

\bibitem[Frayer et al.(2019)]{frayer19} Frayer, D.~T., Maddalena, R.~J., White, S., et al.\ 2019, arXiv e-prints, arXiv:1906.02307

\bibitem[Galametz et al.(2013)]{gala13} Galametz, M., Kennicutt, R.~C., Calzetti, D., et al.\ 2013, \mnras, 431, 1956. doi:10.1093/mnras/stt313


\bibitem[Gao \& Solomon(2004a)]{gs04} Gao, Y., \& Solomon, P.~M.\ 2004, \apj, 606, 271

\bibitem[Gao \& Solomon(2004b)]{gs04b} Gao, Y. \& Solomon, P.~M.\ 2004, \apjs, 152, 63. doi:10.1086/383003

\bibitem[Garcia-Burillo et al.(1993)]{garcia93} Garcia-Burillo, S., Guelin, M., \& Cernicharo, J.\ 1993, \aap, 274, 123

\bibitem[Goldsmith(2001)]{goldsmith01} Goldsmith, P.~F.\ 2001, \apj, 557, 736

\bibitem[Graci{\'a}-Carpio et al.(2006)]{gg06} Graci{\'a}-Carpio, J., Garc{\'\i}a-Burillo, S., Planesas, P., et al.\ 2006, \apjl, 640, L135. doi:10.1086/503361

\bibitem[Graci{\'a}-Carpio et al.(2008)]{gg08} Graci{\'a}-Carpio, J., Garc{\'\i}a-Burillo, S., Planesas, P., et al.\ 2008, \aap, 479, 703

\bibitem[Halfen et al.(2017)]{halfen17} Halfen, D.~T., Woolf, N.~J., \& Ziurys, L.~M.\ 2017, \apj, 845, 158. doi:10.3847/1538-4357/aa816b

\bibitem[Harris et al.(2020)]{harris20} Harris, C.~R., Millman, K.~J., van der Walt, S.~J., et al.\ 2020, \nat, 585, 357. doi:10.1038/s41586-020-2649-2

\bibitem[Haslam et al. (1970)]{haslam70} Haslam, C.~G.~T. and Quigley, M.~J.~S. and Salter, C.~J., \mnras, 147, 405

\bibitem[Heiderman et al.(2010)]{heiderman10} Heiderman, A., Evans, N.~J., Allen, L.~E., et al.\ 2010, \apj, 723, 1019. doi:10.1088/0004-637X/723/2/1019

\bibitem[Hirota et al.(2010)]{hirota10} Hirota, A., Kuno, N., Sato, N., et al.\ 2010, \pasj, 62, 1261. doi:10.1093/pasj/62.5.1261


\bibitem[Iglesias \& Silk(1978)]{ig78} Iglesias, E.~R. \& Silk, J.\ 1978, \apj, 226, 851. doi:10.1086/156665

\bibitem[Imanishi et al.(2006)]{ima06} Imanishi, M., Nakanishi, K., \& Kohno, K.\ 2006, \aj, 131, 2888. doi:10.1086/503527

\bibitem[Imanishi et al.(2007)]{ima07} Imanishi, M., Nakanishi, K., Tamura, Y., et al.\ 2007, \aj, 134, 2366. doi:10.1086/523598

\bibitem[Imanishi et al.(2009)]{ima09} Imanishi, M., Nakanishi, K., Tamura, Y., et al.\ 2009, \aj, 137, 3581. doi:10.1088/0004-6256/137/3/3581


\bibitem[Jim{\'e}nez-Donaire et al.(2017)]{jimnez17} Jim{\'e}nez-Donaire, M.~J., Bigiel, F., Leroy, A.~K., et al.\ 2017, \mnras, 466, 49. doi:10.1093/mnras/stw2996


\bibitem[Jim{\'e}nez-Donaire et al.(2019)]{jimnez19} Jim{\'e}nez-Donaire, M.~J., Bigiel, F., Leroy, A.~K., et al.\ 2019, \apj, 880, 127. doi:10.3847/1538-4357/ab2b95



\bibitem[Kauffmann et al.(2017)]{kauffmann17} Kauffmann, J., Goldsmith, P.~F., Melnick, G., et al.\ 2017, \aap, 605, L5. doi:10.1051/0004-6361/201731123

\bibitem[Kennicutt(1998)]{kennicutt98} Kennicutt, R.~C.\ 1998, \apj, 498, 541. doi:10.1086/305588

\bibitem[Kennicutt et al.(2011)]{kennicutt11} Kennicutt, R.~C., Calzetti, D., Aniano, G., et al.\ 2011, \pasp, 123, 1347. doi:10.1086/663818

\bibitem[Kennicutt, \& Evans(2012)]{ke12} Kennicutt, R.~C., \& Evans, N.~J.\ 2012, \araa, 50, 531

\bibitem[Kreckel et al.(2018)]{kreckel18} Kreckel, K., Faesi, C., Kruijssen, J.~M.~D., et al.\ 2018, \apjl, 863, L21. doi:10.3847/2041-8213/aad77d

\bibitem[Knudsen et al.(2007)]{knudsen07} Knudsen, K.~K., Walter, F., Weiss, A., et al.\ 2007, \apj, 666, 156

\bibitem[Krips et al.(2008)]{krips08} Krips, M., Neri, R., Garc{\'\i}a-Burillo, S., et al.\ 2008, \apj, 677, 262

\bibitem[Krips et al.(2010)]{krips10} Krips, M., Crocker, A.~F., Bureau, M., et al.\ 2010, \mnras, 407, 2261

\bibitem[Kruijssen \& Longmore(2014)]{kruijssen14} Kruijssen, J.~M.~D. \& Longmore, S.~N.\ 2014, \mnras, 439, 3239. doi:10.1093/mnras/stu098

\bibitem[Koda et al.(2009)]{koda09} Koda, J., Scoville, N., Sawada, T., et al.\ 2009, \apjl, 700, L132

\bibitem[Kuno et al.(2007)]{kuno07} Kuno, N., Sato, N., Nakanishi, H., et al.\ 2007, \pasj, 59, 117

\bibitem[Kutner, \& Ulich(1981)]{kutner81} Kutner, M.~L., \& Ulich, B.~L.\ 1981, \apj, 250, 341

\bibitem[Lada \& Lada(2003)]{lada03} Lada, C.~J. \& Lada, E.~A.\ 2003, \araa, 41, 57. doi:10.1146/annurev.astro.41.011802.094844

\bibitem[Lada et al.(2010)]{lada10} Lada, C.~J., Lombardi, M., \& Alves, J.~F.\ 2010, \apj, 724, 687. doi:10.1088/0004-637X/724/1/687

\bibitem[Leroy et al.(2008)]{leroy08} Leroy, A.~K., Walter, F., Brinks, E., et al.\ 2008, \aj, 136, 2782. doi:10.1088/0004-6256/136/6/2782

\bibitem[Leroy et al.(2017)]{leroy17} Leroy, A.~K., Usero, A., Schruba, A., et al.\ 2017, \apj, 835, 217. doi:10.3847/1538-4357/835/2/217

\bibitem[Leroy et al.(2021)]{leroy21} Leroy, A.~K., Schinnerer, E., Hughes, A., et al.\ 2021, \apjs, 257, 43. doi:10.3847/1538-4365/ac17f3





\bibitem[Mangum et al.(2007)]{mangum07} Mangum, J.~G., Emerson, D.~T., \& Greisen, E.~W.\ 2007, \aap, 474, 679

\bibitem[Mangum \& Shirley(2015)]{mangum15} Mangum, J.~G. \& Shirley, Y.~L.\ 2015, \pasp, 127, 266. doi:10.1086/680323




\bibitem[Meier et al.(2000)]{meier00} Meier, D.~S., Turner, J.~L., \& Hurt, R.~L.\ 2000, \apj, 531, 200

\bibitem[Meier \& Turner(2001)]{mt01} Meier, D.~S., \& Turner, J.~L.\ 2001, \apj, 551, 687

\bibitem[Meier \& Turner(2005)]{meier05} Meier, D.~S., \& Turner, J.~L.\ 2005, \apj, 618, 259

\bibitem[Meier et al.(2011)]{meier11} Meier, D.~S., Turner, J.~L., \& Schinnerer, E.\ 2011, \aj, 142, 32

\bibitem[Meijerink et al.(2007)]{meijerink07} Meijerink, R., Spaans, M., \& Israel, F.~P.\ 2007, \aap, 461, 793. doi:10.1051/0004-6361:20066130

\bibitem[Meijerink et al.(2011)]{meij11} Meijerink, R., Spaans, M., Loenen, A.~F., et al.\ 2011, \aap, 525, A119. doi:10.1051/0004-6361/201015136

\bibitem[Momose et al.(2013)]{momose13} Momose, R., Koda, J., Kennicutt, R.~C., et al.\ 2013, \apjl, 772, L13. doi:10.1088/2041-8205/772/1/L13


\bibitem[Murphy et al.(2015)]{murphy15} Murphy, E.~J., Dong, D., Leroy, A.~K., et al.\ 2015, \apj, 813, 118. doi:10.1088/0004-637X/813/2/118



\bibitem[Nguyen-Q-Rieu et al.(1992)]{ng92} Nguyen, Q.-R., Jackson, J.~M., Henkel, C., et al.\ 1992, \apj, 399, 521

\bibitem[Nguyen-Q-Rieu et al.(1994)]{nr94} Nguyen-Rieu, Viallefond, F., Combes, F., et al.\ 1994, IAU Colloq. 140: Astronomy with Millimeter and Submillimeter Wave Interferometry, 336

\bibitem[Pan et al.(2022)]{pan22} Pan, H.-A., Schinnerer, E., Hughes, A., et al.\ 2022, \apj, 927, 9. doi:10.3847/1538-4357/ac474f

\bibitem[Papadopoulos(2007)]{papadopoulos07} Papadopoulos, P.~P.\ 2007, \apj, 656, 792. doi:10.1086/510186


\bibitem[Privon et al.(2015)]{privon15} Privon, G.~C., Herrero-Illana, R., Evans, A.~S., et al.\ 2015, \apj, 814, 39

\bibitem[Privon et al.(2020)]{privon20} Privon, G.~C., Ricci, C., Aalto, S., et al.\ 2020, \apj, 893, 149. doi:10.3847/1538-4357/ab8015



\bibitem[Saha et al.(2002)]{saha02} Saha, A., Claver, J., \& Hoessel, J.~G.\ 2002, \aj, 124, 839. doi:10.1086/341649


\bibitem[Schinnerer et al.(2008)]{sch08} Schinnerer, E., B{\"o}ker, T., Meier, D.~S., et al.\ 2008, \apjl, 684, L21. doi:10.1086/592109

\bibitem[Schruba et al.(2010)]{schruba10} Schruba, A., Leroy, A.~K., Walter, F., et al.\ 2010, \apj, 722, 1699. doi:10.1088/0004-637X/722/2/1699

\bibitem[Schulz et al.(2001)]{schulz01} Schulz, A., G{\"u}sten, R., K{\"o}ster, B., et al.\ 2001, \aap, 371, 25


\bibitem[Sieth et al.(2014)]{sieth14} Sieth, M., Devaraj, K., Voll, P., et al.\ 2014, \procspie, 91530P

\bibitem[Silk(1997)]{silk97} Silk, J.\ 1997, \apj, 481, 703. doi:10.1086/304073

\bibitem[Sheth et al.(2002)]{sheth02} Sheth, K., Vogel, S.~N., Regan, M.~W., et al.\ 2002, \aj, 124, 2581

\bibitem[Shirley(2015)]{shirley15} Shirley, Y.~L.\ 2015, \pasp, 127, 299. doi:10.1086/680342


\bibitem[Tan et al.(2018)]{tan18} Tan, Q.-H., Gao, Y., Zhang, Z.-Y., et al.\ 2018, \apj, 860, 165


\bibitem[Turner \& Ho(1983)]{th83} Turner, J.~L. \& Ho, P.~T.~P.\ 1983, \apjl, 268, L79. doi:10.1086/184033

\bibitem[Turner \& Hurt(1992)]{turner92} Turner, J.~L., \& Hurt, R.~L.\ 1992, \apj, 384, 72


\bibitem[Usero et al.(2015)]{usero15} Usero, A., Leroy, A.~K., Walter, F., et al.\ 2015, \aj, 150, 115. doi:10.1088/0004-6256/150/4/115




\bibitem[Van der Tak et al.(2007)]{vdt07} Van der Tak, F.F.S., Black, J.H., Schöier, F.L., Jansen, D.J., van Dishoeck, E.F., 2007, \aap, 468, 627

\bibitem[Virtanen et al.(2020)]{virtanen20} Virtanen, P., Gommers, R., Oliphant, T.~E., et al.\ 2020, Nature Methods, 17, 261. doi:10.1038/s41592-019-0686-2

\bibitem[Viti(2017)]{viti17} Viti, S.\ 2017, \aap, 607, A118. doi:10.1051/0004-6361/201628877

\bibitem[Wilson(1999)]{wilson99} Wilson, T.~L.\ 1999, Reports on Progress in Physics, 62, 143. doi:10.1088/0034-4885/62/2/002

\bibitem[Yamada et al.(2007)]{yamada07} Yamada, M., Wada, K., \& Tomisaka, K.\ 2007, \apj, 671, 73. doi:10.1086/522332

 
\end{thebibliography}
\end{document}